\documentclass[twocolumn]{openjournal}
\usepackage[utf8]{inputenc}
\usepackage{array}
\usepackage{graphicx}
\graphicspath{images/}
\usepackage[utf8]{inputenc}
\usepackage{xcolor}
\usepackage{enumerate}
\usepackage{graphicx}
\usepackage[caption=false]{subfig}
\usepackage[normalem]{ulem}
\usepackage{amsmath,bm}
\usepackage{hyperref}
\graphicspath{images/}

\begin{document}

\title{Investigating the Dark Energy Constraint from Strongly Lensed AGN at LSST-Scale}
\author{Sydney Erickson}
\affiliation{Kavli Institute for Particle Astrophysics and Cosmology, Department of Physics, Stanford University, Stanford, CA 94309, USA}
\affiliation{SLAC National Accelerator Laboratory, Menlo Park, CA 94025, USA}
\author{Martin Millon} 
\affiliation{Kavli Institute for Particle Astrophysics and Cosmology, Department of Physics, Stanford University, Stanford, CA 94309, USA}
\affiliation{Institute for Particle Physics and Astrophysics, ETH Zurich, CH-8093 Zurich, Switzerland}
\author{Padmavathi Venkatraman}
\affiliation{Department of Astronomy, University of Illinois Urbana-Champaign, Urbana, IL 61801, USA}
\author{Tian Li}
\affiliation{Institute of Cosmology and Gravitation, University of Portsmouth, Portsmouth, UK}
\author{Philip Holloway}
\affiliation{Institute of Cosmology and Gravitation, University of Portsmouth, Portsmouth, UK}
\author{Phil Marshall}
\affiliation{Kavli Institute for Particle Astrophysics and Cosmology, Department of Physics, Stanford University, Stanford, CA 94309, USA}
\affiliation{SLAC National Accelerator Laboratory, Menlo Park, CA 94025, USA}
\author{Anowar J.~Shajib} 
\affiliation{Kavli Institute for Cosmological Physics, University of Chicago, Chicago, IL 60637, USA}
\affiliation{Department of Astronomy and Astrophysics, University of Chicago, Chicago, IL 60637, USA}
\affiliation{Center for Astronomy, Space Science and Astrophysics, Independent University, Bangladesh, Dhaka 1229, Bangladesh}
\author{Simon Birrer}
\affiliation{Department of Physics and Astronomy, Stony Brook University, Stony Brook, NY 11794, USA}
\author{Xiang-Yu Huang} 
\affiliation{Department of Physics and Astronomy, Stony Brook University, Stony Brook, NY 11794, USA}
\author{Timo Anguita}
\affiliation{Instituto de Astrofísica, Facultad de Ciencias Exactas, Universidad Andres Bello, Santiago, Chile}
\author{Steven Dillmann}
\affiliation{Kavli Institute for Particle Astrophysics and Cosmology, Department of Physics, Stanford University, Stanford, CA 94309, USA}
\affiliation{Stanford Artificial Intelligence Laboratory, Department of Computer Science, Stanford University, Stanford, CA 94309, USA}
\affiliation{Institute of Computational and Mathematical Engineering, Stanford University, Stanford, CA 94309, USA}
\author{Narayan Khadka}
\affiliation{Kavli Institute for Particle Astrophysics and Cosmology, Department of Physics, Stanford University, Stanford, CA 94309, USA}
\affiliation{SLAC National Accelerator Laboratory, Menlo Park, CA 94025, USA}
\author{Kate Napier}
\affiliation{Kavli Institute for Particle Astrophysics and Cosmology, Department of Physics, Stanford University, Stanford, CA 94309, USA}
\affiliation{SLAC National Accelerator Laboratory, Menlo Park, CA 94025, USA}
\author{Aaron Roodman} 
\affiliation{Kavli Institute for Particle Astrophysics and Cosmology, Department of Physics, Stanford University, Stanford, CA 94309, USA}
\affiliation{SLAC National Accelerator Laboratory, Menlo Park, CA 94025, USA}
\author{the LSST Dark Energy Science Collaboration}
\noaffiliation
\begin{abstract}

Strongly lensed Active Galactic Nuclei (AGN) with an observable time delay can be used to constrain the expansion history of the Universe through time-delay cosmography (TDC). As the sample of time-delay lenses grows to statistical size, with $\mathcal{O}$(1000) lensed AGN forecast to be observed by the Vera C. Rubin Observatory Legacy Survey of Space and Time (LSST), there is an emerging opportunity to use TDC as an independent probe of dark energy. To take advantage of this statistical sample, we implement a scalable hierarchical inference tool which computes the cosmological likelihood for hundreds of strong lenses simultaneously. With this new technique, we investigate the cosmological constraining power from a simulation of the full LSST sample. We start from individual lenses, and emulate the full joint hierarchical TDC analysis, including image-based modeling, time-delay measurement, velocity dispersion measurement, and external convergence prediction. We fully account for the mass-sheet and mass-anisotropy degeneracies. We assume a sample of 800 lenses, with varying levels of follow-up fidelity based on existing campaigns. With our baseline assumptions, within a flexible $w_0w_a$CDM cosmology, we simultaneously forecast a $\sim$2.5\% constraint on $H_0$ and a dark energy figure of merit (DE FOM) of 6.7. We show that by expanding the sample from 50 lenses with IFU kinematics to include an additional 750 lenses with plausible LSST time-delay measurements, we improve the forecasted DE FOM by nearly a factor of 3, demonstrating the value of incorporating this portion of the sample. We also investigate different follow-up campaign strategies, and find significant improvements in the DE FOM with additional stellar kinematics measurements and higher-precision time-delay measurements. We also demonstrate how the redshift configuration of time-delay lenses impacts constraining power in $w_0w_a$CDM. 

\end{abstract}

\section{Introduction}
\defcitealias{TDCOSMO_2025}{TDCOSMO25}

As our ability to measure the effect of dark energy (DE) on cosmic expansion history improves, there is growing observational evidence for time-evolution of the DE equation of state \citep{DESI_DR2_cosmology}. This potential dynamical nature of DE is of high interest, given unresolved challenges associated with the cosmological constant model \citep{weinberg89, velten_2014}. In this landscape, time-delay cosmography (TDC) contributes as an independent probe of cosmological expansion, providing a single-step angular diameter distance constraint in the late Universe \citep[see e.g.][]{tdc_review_24}. The DE equation of state has recently been constrained from the TDCOSMO sample of eight lensed quasars (\citet{TDCOSMO_2025}, hereafter \citetalias{TDCOSMO_2025}), demonstrating the cosmological information present in these systems \citep[also shown by][]{Hogg_2023,shajib_frieman_25}. Although TDC already yields a 4$\%$ measurement of the Hubble constant, $H_0$, its constraining power on the DE equation of state parameters, ($w_0$, $w_a$), is currently weak. However, the Vera C. Rubin Observatory Legacy Survey of Space and Time (LSST) will expand the time-delay lens sample by 1-2 orders of magnitude, creating a promising opportunity to better constrain DE through TDC \citep{coe_2009, om10, shajib_SLSC}. In addition, through the combination of a larger sample of strongly lensed Active Galactic Nuclei (AGNs) with other strong lensing systems, including strongly lensed supernovae and galaxy-galaxy strong lenses, the joint strong lensing probe is emerging as a competitive option for constraining the DE equation of state \citep{shajib_SLSC}. Given the potential for TDC as a DE probe, we aim to further investigate the projected constraint from LSST lensed AGN. 

The DE measurement from a statistical sample of lensed AGN will be made feasible by LSST. The combination of unprecedented depth and time-domain information for nearly the entire Southern sky makes the LSST a powerful discovery tool for new lensed AGN. A combination of image-based and time-domain-based finding methods will allow us to discover thousands of new lensed AGN \citep{finding_review}. The LSST survey is forecasted to contain several thousand observable lensed AGN, with estimates ranging from $\sim$2400 to $\sim$3500 systems in total \citep{om10, yue_2022, abe_2025}. Additionally, it is predicted that $\sim$30-40\% of the sample will have variability that is detectable by LSST \citep{taak_treu}. Given these forecasts, we assume a final sample size of 800 lenses with time-delays measurable by LSST, increasing the sample size to be $\sim$4 times larger than what has been used in recent TDC forecasts \citep{tdcosmo_V, shajib_SLSC}.

To cope with the increase in sample size, the community has been working towards scalable modeling routines, with open challenges for image modeling \citep{tdlmc_2021} and time-delay extraction from LSST light curves \citep{timedelay_challenge}. 
In this work, we turn our attention to the population inference step of the analysis. In TDC, we use a hierarchical Bayesian framework to combine the constraints from all lenses, which share an informative prior for some of their parameters. This method was developed in \citet{tdcosmo_IV} and \citetalias{TDCOSMO_2025}. This existing hierarchical inference method uses the \textsc{hierArc}\footnote{\url{https://github.com/sibirrer/hierArc}} code. In this work, we present a new likelihood evaluation code, \textsc{fasttdc}\footnote{\url{https://github.com/smericks/fasttdc}}, that is optimized for larger samples of lenses. We compute a set of static data vectors summarizing the models and measurements of each individual lens. Then, we feed the data vectors to a hierarchical Bayesian inference, where the cosmological likelihood is evaluated over all lenses simultaneously. Data vector quantities are chosen such that they do not depend on the parameterization of individual lens models. 

With this newly developed framework, we enable investigation of the predicted cosmological constraint from a simulated sample of 800 time-delay lenses. We start from lenses in the OM10 catalog  \citep{om10}, and emulate each portion of the modeling and observation process. We then process the sample through a joint hierarchical Bayesian inference for cosmological parameters and lens population properties, fully accounting for the mass-sheet and mass-anisotropy degeneracies. In addition to a baseline forecast, we also investigate how additional follow-up campaigns, and redshift configuration  impact the final constraint. As our primary metric, we adopt the Dark Energy Task Force figure of merit (DE FOM) \citep{DETF}. We aim to answer the following questions:
\begin{itemize}

    \item How much cosmological constraining power is contained within the larger sample of LSST lenses when combined with the smaller, more extensively studied time-delay lens sample? 

    \item How does the DE FOM depend on the portion of the lens sample with stellar kinematics? Should we use telescope time to measure stellar kinematics for many lenses in a single aperture, or fewer lenses with spatially resolved kinematics, using Integral Field Units (IFU)?

    \item How sensitive is the DE FOM to mass models from high-resolution imaging versus less precise mass models, obtained either from ground-based data and/or automated modeling? What about the impact of long-term time-delay monitoring, versus time-delay measurement from LSST light curves?

    \item Which redshift configurations are most advantageous for measuring dark energy from time-delay lenses?     

\end{itemize}

These questions are important to answer as the LSST begins so we can optimize the follow-up campaigns and modeling efforts the community will embark to build towards the first Dark Energy Science Collaboration (DESC) TDC cosmology result. In Section \ref{section:background}, we provide background on how TDC can be used to measure cosmological expansion. In Section \ref{section:method}, we describe our hierarchical inference method. In Section \ref{section:lens_sample}, we describe how we emulate measurements for a simulated lensed AGN sample. In Section \ref{section:experiments}, we describe our full set of experiments, and provide results. In Sections \ref{section:discussion} and \ref{section:conclusion}, we discuss our results and give final conclusions.

\section{Background}
\label{section:background}

Strongly lensed AGN contain cosmological information through their time-delays (Section \ref{subsection:tdc}) and the relation of the lens deflection in angular units to the absolute mass of the lens, measured by stellar kinematics (Section \ref{subsection:kin}). Using both time-delays and stellar kinematics breaks important lensing degeneracies (Section \ref{subsection:degen}). From a population of lenses, a hierarchical inference framework is used to infer the cosmology in which all of the lenses reside (Section \ref{subsection:HI}). 

\subsection{Time Delay Cosmography}
\label{subsection:tdc}

The differences in the arrival time of light to the observer between the multiple images in a strongly lensed AGN depend on the mass and geometry of the lens \citep{refsdal64}. This relationship is explained by the time delay equation:
\begin{equation}
    \Delta t_{\text{AB}} = \frac{1}{c} D_{\Delta t} \Delta \phi_{AB}(\xi_{\text{lens}}).
\label{eqn:time_delay}
\end{equation}
The measured time delay between image A and image B, $\Delta t_{\text{AB}}$, is related to the difference in the Fermat potential at each image position, $\Delta \phi_{AB}(\xi_{\text{lens}})$. The mass model of the lens, parameterized by $\xi_{\text{lens}}$, determines the Fermat potential. The geometry of the source-lens-observer system enters through the time-delay distance:
\begin{equation}
    D_{\Delta t} \equiv (1+z_{\text{lens}})\frac{D_{\text{d}} D_{\text{s}}}{D_{\text{ds}}}.
\end{equation}
The time-delay distance is a ratio of angular diameter distances, where $D_{\text{d}}$ is the distance to the deflector, $D_{\text{s}}$ is the distance to the source, and $D_{\text{ds}}$ is the distance between the deflector and the source. These distances are a function of both redshift and cosmological parameters, including $H_0$ and others controlling the expansion history ($\Omega_{\text{m}}$, $w_0$, and $w_a$ in $w_0w_a$CDM). In particular, we are interested in constraining the DE equation of state, which describes the relationship between pressure, $p$ and energy density, $\rho$, as: $w = p / \rho$ \citep[see e.g.][]{DE_review}. In the cosmological constant model, $w=-1$. In $w_0w_a$CDM, the equation of state follows the parameterization from \citet{chevallier_polarski} and \citet{linder2003}: 
\begin{equation}
    w(a) = w_0 + w_a(1-a),
\end{equation}
where $a$ is the scale factor. Using $D_{\Delta t}$ to constrain cosmological parameters can be thought of as a distance-redshift test, where angular diameter distances are anchored by the time-delay constraint through Equation \ref{eqn:time_delay}.

\subsection{Lensing + Stellar Kinematics}
\label{subsection:kin}

The angular mass profile of a lensing galaxy is related to the absolute mass of the galaxy through the line-of-sight projected stellar velocity dispersion:

\begin{equation}
    \sigma_v = \sqrt{\frac{D_{\text{s}}}{D_{\text{ds}}} c^2 \mathcal{J}(\xi_{\text{lens}})}.
\label{eqn:kinematics}
\end{equation}

The measured line-of-sight projected stellar velocity dispersion, $\sigma_v$, is a function of both the cosmology-independent portion of a spherical Jeans model of the lens, $\mathcal{J}(\xi_{\text{lens}})$, and a ratio of angular diameter distances, $\frac{D_{\text{s}}}{D_{\text{ds}}}$ \citep{Birrer16_v_disp_eqn, Birrer19_H0LiCOW_IX}. Here, $\xi_{\text{lens}}$ are the parameters that describe a lens model in angular units, fit from imaging data. With a predicted $\mathcal{J}(\xi_{\text{lens}})$ from image-based lens modeling, an observed velocity dispersion, $\sigma_v$, and redshifts of the deflector and source, one can constrain cosmology through the ratio of angular diameter distances. Note that this probe, excluding the time-delay information, is \textit{not} sensitive to the Hubble constant, but is still sensitive to other cosmological parameters. We present the time-delay and stellar kinematics constraints separately, but in practice, the two probes are used jointly in time-delay cosmography, to account for important degeneracies described below.

\subsection{Degeneracies}
\label{subsection:degen}

The relationships established in Equations 
\ref{eqn:time_delay} and \ref{eqn:kinematics} are additionally sensitive to degeneracies. The two effects are mass-sheet degeneracy \citep{falco_1985} and mass-anisotropy degeneracy \citep{binney_1982}.

Mass-sheet degeneracy explains that if one were to add a constant and infinite sheet of mass to the convergence profile of a lens, while simultaneously scaling the convergence profile, the angular model of the lens deflection would not change, but the measured time-delay and velocity dispersion would \citep{schneider_sluse_2013}. In this work, we consider both an internal mass sheet, $\lambda_{\text{int}}$, and an external mass sheet, described as an external convergence, $\kappa_{\text{ext}}$. External convergence can be constrained through additional line-of-sight (LOS) observables, such as galaxy number counts \citep[see e.g.][]{ext_conv_galaxy_counts}. We note that a more complex treatment of the foreground LOS environment, and the potential bias it may induce on TDC measurements \citep{johnson_LOS}, will be investigated in future work. The internal mass sheet, $\lambda_{\text{int}}$, is only constrained by combining both the time-delay and kinematic observables.

Mass-anisotropy degeneracy introduces the effect of anisotropic stellar orbits. Anisotropic orbits change the predicted Jeans model $\mathcal{J}(\xi_{\text{lens}})$ $\rightarrow$ $\mathcal{J}(\xi_{\text{lens}},\rm \beta_{ani} )$, used in Equation \ref{eqn:kinematics}. Each lensing galaxy is characterized by an anisotropy parameter, $\beta_{\text{ani}} = 1 - \sigma_t^2 / \sigma_r^2$, which describes the ratio between tangential and radial velocity dispersions ($\sigma_t^2$ and $\sigma_r^2$).  We assume a constant anisotropy profile, matching the assumption made in \citetalias{TDCOSMO_2025}. This parameter is only constrained through multiple measurements of the velocity dispersion at different radii, so it is crucial that some lenses have spatially-resolved kinematic measurements \citep{SAURON_X}.

Accounting for mass-sheet degeneracy and mass-anisotropy degeneracy, the time-delay and velocity dispersion equations (Equations \ref{eqn:time_delay} and \ref{eqn:kinematics}) become: 

\begin{equation}
    \Delta t_{\text{AB}} = \lambda_{\text{int}} (1-\kappa_{\text{ext}}) \frac{1}{c} D_{\Delta t} \Delta \phi_{AB}(\xi_{\text{lens}}),
\label{eqn:time_delay_degen}
\end{equation}

and 
\begin{equation}
    \sigma_v = \rm \sqrt{\lambda_{\text{int}} (1-\kappa_{\text{ext}}) \frac{D_{\text{s}}}{D_{\text{ds}}} c^2 \mathcal{J}(\xi_{\text{lens}},\beta_{\text{ani}})}.
\label{eqn:kinematics_degen}
\end{equation}

In the rest of this work, we use the relations established in Equations \ref{eqn:time_delay_degen} and \ref{eqn:kinematics_degen}.

\subsection{Joint Hierarchical Inference}
\label{subsection:HI}

Building upon the cosmological constraining power contained in individual lenses, we exploit the self-similarity of lenses by using the joint hierarchical Bayesian inference framework laid out in \cite{tdcosmo_IV}. At the population level, we jointly infer $w_0w_a$CDM cosmological parameters:

\begin{equation}
    \Omega = \{H_0,\Omega_{\text{m}},w_0,w_a\},
\end{equation}
and lens galaxy population properties
\begin{equation}
    \nu = \{\mu(\lambda_{\text{int}}),\sigma(\lambda_{\text{int}}),\mu(\beta_{\text{ani}}),\sigma(\beta_{\text{ani}}) \}.
\end{equation}

We assume all of the lenses share a population distribution in $\lambda_{\text{int}}$ and $\beta_{\text{ani}}$. We assume those populations are Gaussian, resulting in the mean and standard deviation hyperparameters in $\nu$ that describe the lens galaxy population model.

The inference is informed by an individual dataset for each lens:
\begin{equation}
    \mathcal{D}_k = \{d_{\text{img}}, d_{\text{td}}, d_{\text{kin}}, d_{\text{los}}\}.
\end{equation}

Every lens has $d_{\text{img}}$, the image data used for mass modeling, $d_{\text{td}}$, the time-delay measurements, derived from light-curves, and $d_{\text{los}}$, the LOS external convergence measurements. Some lenses additionally have velocity dispersion measurements derived from spectroscopic data, $d_{\text{kin}}$.

We start with Bayes' proportionality relation:

\begin{equation}
    p(\Omega,\nu|\mathcal{D}) \propto p(\mathcal{D}|\Omega,\nu)p(\Omega,\nu),
\label{eqn:bayes}
\end{equation}

where the posterior over the target quantities ($\Omega,\nu$) given the data ($\mathcal{D}$) is proportional to the product of the likelihood, $p(\mathcal{D}|\Omega,\nu)$, and the prior, $p(\Omega,\nu)$. We assume that each lens provides an independent constraint: 
\begin{equation}
    p(\Omega,\nu|\mathcal{D}) \propto p(\Omega,\nu) \prod_k p(\mathcal{D}_k|\Omega,\nu).
\label{eqn:posterior_relation}
\end{equation}

The challenge here is evaluating the likelihood for each lens, $p(\mathcal{D}_k|\Omega,\nu)$. First, we expand the evaluation, taking the product of the likelihood for each observable. Then, we introduce a marginalization over individual lens properties. We include a full likelihood derivation in Appendix \ref{appendix:likelihood_deriv}. The likelihood of an individual lens becomes: 
\begin{align}
p(\mathcal{D}_k|\Omega,\nu) \propto \int &\, 
    p(d_{\text{td}} \mid \Omega, \lambda_{\text{int}}, \kappa_{\text{ext}}, \Delta\phi(\xi_{\text{lens}})) \nonumber \\
    &\times p(d_{\text{kin}} \mid \Omega, \lambda_{\text{int}}, \kappa_{\text{ext}}, \mathcal{J}(\xi_{\text{lens}},\beta_{\text{ani}})) \nonumber \\
    &\times p(d_{\text{img}} \mid \xi_{\text{lens}}) \nonumber p(d_{\text{los}} \mid \kappa_{\text{ext}}) \nonumber \\
    &\times p(\lambda_{\text{int}}, \beta_{\text{ani}} \mid \nu)  p(\kappa_{\text{ext}}, \xi_{\text{lens}}) \nonumber \\
    & \mathrm{d}\lambda_{\text{int}}\, \mathrm{d}\kappa_{\text{ext}}\, \mathrm{d}\beta_{\text{ani}}\, \mathrm{d}\xi_{\text{lens}}.
\label{eqn:full_likelihood}
\end{align}
Note the marginalization over individual lens properties: $\lambda_{\text{int}}$, $\kappa_{\text{ext}}$, $\beta_{\text{ani}}$, $\xi_{\text{lens}}$ . Lens and source redshifts are not marginalized over, as we assume perfect knowledge of the redshift for every lens. This equates to assuming some spectroscopic follow-up for the lens and source redshift for every lens in the sample of 800, which is feasible given dedicated surveys like Chilean AGN/Galaxy Extragalactic Survey (ChANGES) \citep{CHANGES_survey}. We discuss spectroscopic campaigns further in Section \ref{subsection:stellar_kin}. 

\begin{figure*}[hbt!]
    \centering
    \includegraphics[scale=0.5]{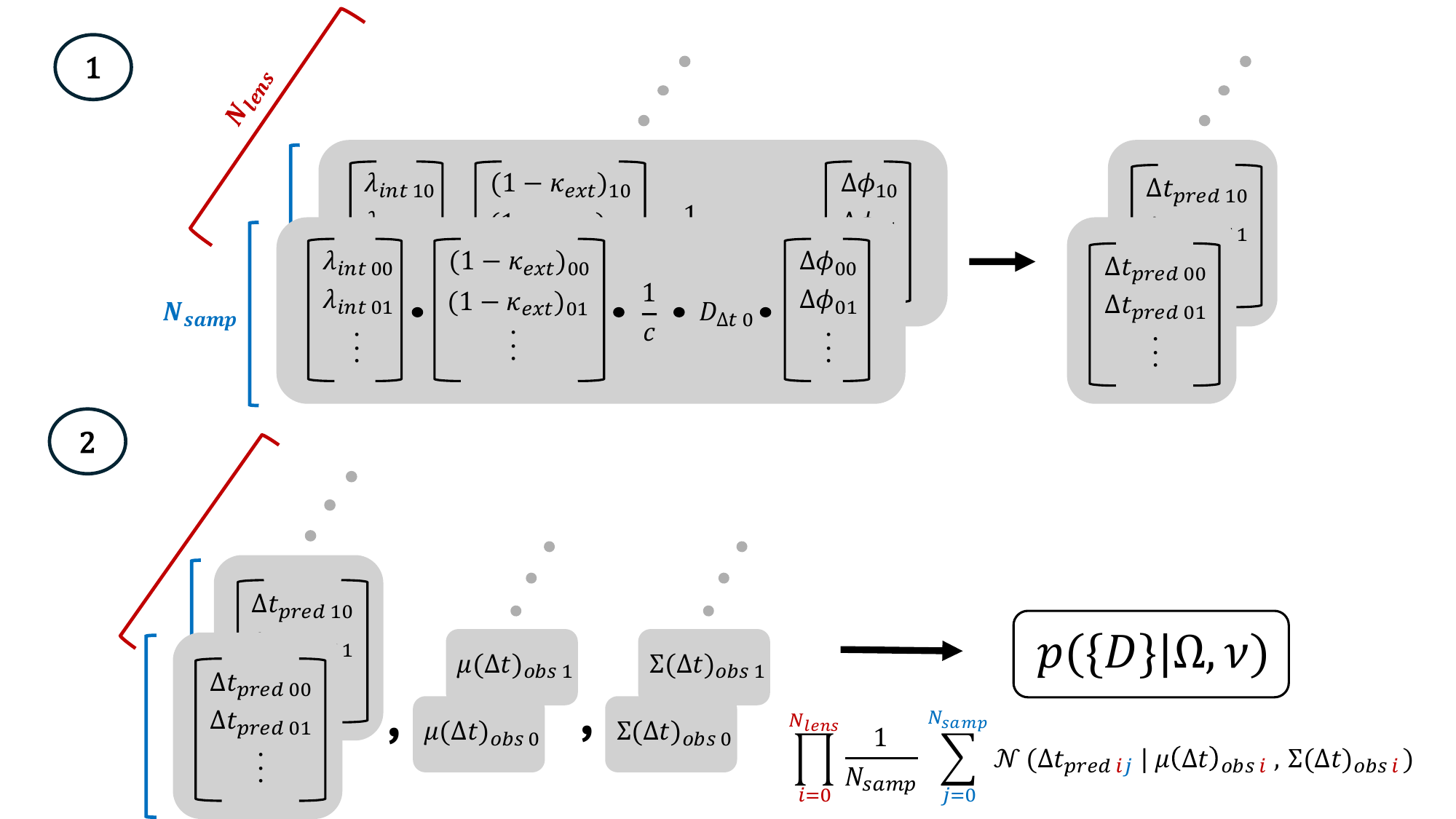}
    \caption{Diagram of the vectorized likelihood evaluation, isolated to the time-delay likelihood only (the kinematic likelihood is treated similarly). In step 1, samples of the modeling inputs ($\lambda_{\text{int}}$,$\kappa_{\text{ext}}$,$\Delta \phi$) are condensed into samples of the predicted time-delay, with a value tracked for every importance sample, across every lens. In step 2, the likelihood of the predicted time-delay is evaluated against the observed time-delay over every importance sample, averaged, and then multiplied across every lens, to produce the final likelihood. See Equation \ref{eqn:imp_sampling_lklhd}.}
    \label{fig:likelihood_diagram}
\end{figure*}

\section{Inference Method}
\label{section:method}

As we prepare for the increase in sample size of time-delay lenses, the accompanying hierarchical inference technique needs to be scalable. We divide the analysis into two stages. In stage 1, we summarize individual lens models and measurements with a static set of data vectors. In stage 2, these data vectors are fed into the hierarchical cosmological inference. We require the data vectors to be agnostic to upstream modeling choices, retaining as much flexibility as possible. The likelihood evaluation is performed simultaneously over many lenses, using vectorized operations to improve the future scalability of the method.

\subsection{Likelihood Evaluation}
\label{subsection:likelihood_eval}

We design a likelihood evaluation code, \textsc{fasttdc}, that operates over hundreds of lenses simultaneously, which we achieve through the vectorized approach summarized in Figure \ref{fig:likelihood_diagram}.

We further modify the likelihood evaluation, starting from Equation \ref{eqn:full_likelihood}. First, we exchange likelihoods for posteriors where appropriate, using Equation \ref{eqn:lklhd_to_posterior_prop} derived in Appendix \ref{subsection:supporting_deriv} to account for the dependence on interim modeling priors, $\rm \nu_{int}$. Full details of this step of the derivation are found in Appendix \ref{appendix:likelihood_deriv}. The likelihood evaluation becomes:
\begin{align}
p(\mathcal{D}_k|\Omega,\nu) \propto &  \int 
    p(d_{\text{td}} | \Omega, \lambda_{\text{int}}, \kappa_{\text{ext}}, \Delta\phi(\xi_{\text{lens}})) \notag \\
    &  \times p(d_{\text{kin}} | \Omega, \lambda_{\text{int}}, \kappa_{\text{ext}}, \mathcal{J}(\xi_{\text{lens}},\beta_{\text{ani}}))) \notag \\
    &  \times p(\xi_{\text{lens}}|d_{\text{img}},\nu_{\text{int}}) \notag \\
    &  \times p(\kappa_{\text{ext}} | d_{\text{los}}, \nu_{\text{int}}) p(\lambda_{\text{int}}, \beta_{\text{ani}} | \nu) \notag \\
    & \rm d\lambda_{\text{int}}d\kappa_{\text{ext}}d\beta_{\text{ani}}d\xi_{\text{lens}}.
\label{eqn:lklhd_integral_full}
\end{align}
Note the introduction of a re-weighting term to account for the informative modeling prior on $\beta_{\text{ani}}$, $\nu_{\text{int}}$. We use importance sampling to evaluate the integral, resulting in the likelihood evaluation:
\begin{align}
    \nonumber p(\mathcal{D}_k \mid \Omega, \nu) \propto 
    \frac{1}{N} & \sum_{\substack{
        \xi_{\text{lens}},\, \beta_{\text{ani}}, \lambda_{\text{int}}, \kappa_{\text{ext}}  \sim \\
        p(\xi_{\text{lens}} \mid d_{\text{img}}, \nu_{\text{int}}) p(\beta_{\text{ani}} | \nu_{\text{int}}) \\
        p(\kappa_{\text{ext}} \mid d_{\text{los}}) p(\lambda_{\text{int}} \mid \nu)
    }} 
     \Bigl[ \\
     \nonumber & \quad p(d_{\mathrm{td}} \mid \Omega, \lambda_{\mathrm{\text{int}}}, \kappa_{\mathrm{\text{ext}}}, \Delta\phi(\xi_{\text{lens}})) \\
     \nonumber & \quad \times\, p(d_{\mathrm{kin}} \mid \Omega, \lambda_{\mathrm{\text{int}}}, \kappa_{\mathrm{\text{ext}}}, \mathcal{J}(\xi_{\text{lens}}, \beta_{\text{ani}})) \\
     & \quad \times\, p(\beta_{\text{ani}} \mid \nu) / p(\beta_{\text{ani}} \mid \nu_{\text{int}}) 
    \Bigr].
\label{eqn:imp_sampling_lklhd}
\end{align}

We demonstrate this computation in Figure \ref{fig:likelihood_diagram}. There are several advantages to this formulation. We do not condense the Fermat potential and the time-delay into a $D_{\Delta t}$ posterior before cosmological likelihood evaluation. Keeping the quantities separated is crucial in order to track the correlation between the predicted Fermat potentials and the velocity dispersion, (as shown in \citet{wang_GLAD}, \citetalias{TDCOSMO_2025}). If a kinematic model provides extra constraining power on the mass model, the Fermat potentials are able to be pulled in the right direction since the correlation is tracked. This is especially relevant for the power-law slope (see Figure B.1 in \citetalias{TDCOSMO_2025}). Note that $\beta_{\text{ani}}$, which enters through the Jeans model $\mathcal{J}(\xi_{\text{lens}},\rm \beta_{ani} )$, has a non-linear relationship to the predicted velocity dispersion. Since the relationship between $\beta_{\text{ani}}$ and $\mathcal{J}$ is tracked sample to sample, this non-linear relationship can be accounted for in this framework.

Given this design for the likelihood evaluation, the static quantities we need from analysis of each lens are as follows. We need samples from the modeling posteriors and their associated pre-computed model quantities:
\begin{itemize}
    \item $ \xi_{\text{lens}},\beta_{\mathrm{ani}} \sim p(\xi_{\text{lens}} | d_{\text{img}}, \nu_{\text{int}}) p(\beta_{\text{ani}} | \nu_{\text{int}})$
    \item $\Delta\phi(\xi_{\text{lens}}) , \mathcal{J}(\xi_{\text{lens}}, \beta_{\mathrm{ani}})$ 
    \item $\kappa_{\text{ext}} \sim p(\kappa_{\text{ext}} | d_{\text{los}}) $,
\end{itemize}

and measurements with Gaussian uncertainties:
\begin{itemize}
    \item $\mu_{\text{obs}}(\mathbf{\Delta t})$, $\Sigma_{\text{obs}}(\mathbf{\Delta t})$
    \item $\mu_{\text{obs}}(\bm{\sigma_v})$, $\Sigma_{\text{obs}}(\bm{\sigma_v})$.
\end{itemize}
All of the time-delays between independent pairs of images are contained within the vector, $\mathbf{\Delta t}$, with covariances $\Sigma_{\text{obs}}(\mathbf{\Delta t})$. Similarly, $\bm{\sigma_v}$ contains all velocity dispersion measurements in a single aperture or multiple radial apertures, with uncertainties stored in the covariance matrix $\Sigma_{\text{obs}}(\bm{\sigma_v})$.

\subsection{Hierarchical Inference}
\label{subsection:HI_method}

The hierarchical inference stage infers the population model posterior from the individual lens models and measurements (Equation \ref{eqn:posterior_relation}). We infer a posterior over eight parameters: $ \Omega,\nu = \{ H_0, \Omega_{\text{m}}, w_0, w_a, \mu(\lambda_{\text{int}}), \sigma(\lambda_{\text{int}}), \mu(\beta_{\text{ani}}), \sigma(\beta_{\text{ani}}) \}$, given the priors detailed in Table \ref{tab:sampling_prior}. We use a sampling approach, employing Markov Chain Monte Carlo (MCMC) with the \textsc{emcee}\footnote{\url{https://github.com/dfm/emcee}} sampler \citep{emcee}. We use an ensemble sampler with 50 walkers, allowing at least five walkers for each free parameter. As the sampler explores the parameter space, the posterior is evaluated many times. For each evaluation, the likelihood over all lenses must be computed, which requires computing hundreds of the individual likelihoods given by Equation \ref{eqn:imp_sampling_lklhd}. Each individual likelihood requires an integration over lens properties. We use 5000 importance samples over individual lens properties to numerically evaluate each integral. For a sample of 800 lenses, this results in each MCMC step taking $\sim$20 seconds. To speed up the run-time, we use MPI to parallelize the sampling of MCMC chains over multiple CPUs. We run the chains for 70k steps, which takes roughly 90 hours when parallelized over 32 CPUs, for a total of $\sim$3100 CPU hours per hierarchical inference run. We use the Stanford University Sherlock computing cluster for all runs. 

\begin{table}[hbt]
\centering
\begin{tabular}{| c | c |}
    \hline
    $H_0$ & $\mathcal{U}(0,150)$\\ 
    \hline
    $\Omega_{\text{m}}$ & $\mathcal{U}(0.05,0.5)$\\ 
    \hline
    $w_0$ & $\mathcal{U}(-2,0)$\\
    \hline
    $w_a$ & $\mathcal{U}(-2,2)$\\
    \hline
    $\mu(\lambda_{\text{int}})$ & $\mathcal{U}(0.5,1.5)$\\
    \hline
    $\sigma(\lambda_{\text{int}})$ & $\mathcal{U}(0.001,0.5)$\\
    \hline
    $\mu(\beta_{\text{ani}})$ & $\mathcal{U}(-0.5,0.5)$\\
    \hline
    $\sigma(\beta_{\text{ani}})$ & $\mathcal{U}(0.001,0.2)$\\
    \hline
\end{tabular}
\caption{Prior over hyperparameters used during Bayesian inference. This corresponds to  $p(\Omega,\nu)$ in Equation \ref{eqn:bayes}.}
\label{tab:sampling_prior}
\end{table}

\section{Simulation of Data Vectors}
\label{section:lens_sample}

In this section, we detail how we generate data vectors from a catalog of lensed quasars in LSST. We start from a simulated lens catalog (Section \ref{subsection_OM10}). Then, we emulate the modeling and measurement of those lenses. We emulate the following products for each lens: the modeling posterior from image data (Section \ref{subsection:image_models}), the time-delay measurement (Section \ref{subsection:time_delays}), the kinematic measurement (Section \ref{subsection:stellar_kin}), and the external convergence posterior from LOS data (Section \ref{subsection:ext_conv}). In this work, we assume a ground truth $\Lambda$CDM model with: $H_0$ = 70 km s$^{-1}$ Mpc$^{-1}$, $\Omega_{\text{m}}$ = 0.3, $w_0$ = -1, $w_a$ = 0.

\subsection{Simulated Lensed AGN Sample}
\label{subsection_OM10}

We start from a simulation of the LSST lensed AGN sample, using the OM10 catalog \citep{om10}, which contains $\sim$3100 lensed AGN with lens and source properties. Following the preparation of \cite{venkatraman_2025}, we modify the catalog to assign a power-law slope to each lens, relaxing the assumption of an isothermal profile for every lens. We also assign each lens an underlying mass-sheet parameter, $\lambda_{\text{int}}$, and an anisotropy parameter, $\beta_{\text{ani}}$. We assume Gaussian populations in both parameters, with $\lambda_{\text{int}} \sim \mathcal{N}(\mu=1,\sigma=0.1)$ and $\beta_{\text{ani}} \sim \mathcal{N}(\mu=0,\sigma=0.1)$, based off results from \citetalias{TDCOSMO_2025}. By assuming lenses have intrinsic scatter in their mass-sheet and anisotropy properties, we make a more conservative choice when forecasting constraining power. When the underlying lenses are more self-similar, the joint inference is more constraining.

\begin{table*}[hbt]
\centering
\begin{tabular}{| c | c | c | c | c | c | c |}
    \hline
    
    \textbf{Lens Type} & \# Lenses & Kinematic Type & $\sigma_v$ Precision & $\Delta t$ Precision & Image Model Type & Median $\Delta \phi$ Precision\\
    \hline \hline

    \multicolumn{7}{|c|}{\textbf{Platinum: Spatially Resolved Kinematics, Dedicated $\Delta t$ Monitoring, High. Res. Imaging}} \\
    \hline
    Platinum-JWST & 10 & NIRSpec & 5\% & 3\% & JWST-FM & 2\% \\
    Platinum-VLT & 40 & MUSE & 5\% & 3\% & HST-FM & 4\% \\
    \hline 

    \multicolumn{7}{|c|}{\textbf{Gold: Single-Aperture Kinematics, LSST $\Delta t$ Monitoring, High. Res. Imaging}} \\
    \hline
    Gold-4MOST & 150 & 4MOST/Magellan & 5\% & 5 days & HST-NPE & 11\% \\
    \hline 
    \multicolumn{7}{|c|}{\textbf{Silver: Single-Aperture / No Kinematics, LSST $\Delta t$ Monitoring, LSST Imaging}} \\
    \hline
    Silver-4MOST & 300 & 4MOST/Magellan & 5\% & 5 days & LSST-NPE & 18\% \\
    Silver & 300 & None & - & 5 days & LSST-NPE & 18\% \\
    \hline
\end{tabular}
\caption{Assumed lens sample in the baseline experiment. We separate the lens sample into three main categories: Platinum, Gold, and Silver. We note the assumed modeling fidelity for each lens type. }
\label{table:baseline}
\end{table*}

\subsection{Image-Based Lens Models}
\label{subsection:image_models}

Once we have produced a catalog of lenses, we emulate the modeling and measurement for every lens. First, we emulate the image-based modeling portion of the analysis. We emulate image-based mass models, $p(\xi_{\text{lens}}|d_{\text{img}},\nu_{\text{int}})$, by applying automated modeling to simulated images of each lens. Then, we store computations of relevant mass-model derived quantities: the Fermat potential differences, $\Delta \phi(\xi_{\text{lens}})$, and Jeans model quantities, $\mathcal{J}(\xi_{\text{lens}},\beta_{\text{ani}})$, for many samples from that mass model posterior.  

To produce mass models from simulated images, we start by simulating both Hubble Space Telescope (HST) and LSST quality images for every lens using the \textsc{paltas} simulation tools, which are based on the \textsc{lenstronomy}\footnote{\url{https://github.com/lenstronomy/lenstronomy}} package \citep{birrer2018lenstronomy,birrer2021lenstronomy}. For HST images, we simulate a 1400s exposure in the F814W filter, following \cite{Erickson_2025}. For LSST images, we simulate a 5 year co-add in the \textit{i}-band, following \cite{venkatraman_2025}. 

Then, we apply a fast machine-learning based modeling technique to the images to produce a Gaussian estimate of the mass model posterior. We apply the \textsc{paltas}\footnote{\url{https://github.com/swagnercarena/paltas}} neural posterior estimation (NPE) technique for strong lens modeling \citep{paltas_paper}, using the training configuration developed in \cite{Erickson_2025} for strongly lensed AGN. We apply the trained neural network to each image to produce an approximate Gaussian posterior $p(\xi_{\text{lens}}|d_{\text{img}},\nu_{\text{int}})$. 

Given mass models for every lens, the next step is to pre-compute the quantities $\Delta \phi(\xi_{\text{lens}})$ and $\mathcal{J}(\xi_{\text{lens}},\beta_{\text{ani}})$ associated with every importance sample $\xi_{\text{lens}} \sim p(\xi_{\text{lens}}|d_{\text{img}},\nu_{\text{int}})$. This step is necessary because the evaluation of $\mathcal{J}(\xi_{\text{lens}},\beta_{\text{ani}})$ is much more computationally intensive than the rest of the cosmological likelihood evaluation. Encountering this computational cost during data vector generation, rather than during likelihood evaluation, allows faster iterations at the cosmological inference stage. See the Derivation in Appendix \ref{appendix:likelihood_deriv} for further details.

First, we generate samples from the posterior, $p(\xi_{\text{lens}}|d_{\text{img}},\nu_{\text{int}})$, and compute the Fermat potential associated with each sample. From these samples, we fit a new multivariate Gaussian to accommodate the $\Delta\phi$ dimension(s), giving: $p(\xi_{\text{lens}}|d_{\text{img}},\nu_{\text{int}})$ $\rightarrow$ $p(\Delta\phi,\xi_{\text{lens}}|d_{\text{img}},\nu_{\text{int}})$. In practice, this is not a true posterior over both ($\Delta\phi,\xi_{\text{lens}}$), but rather a Gaussian model to describe how samples from the lens model posterior translate to $\Delta\phi$ values for each individual lens. We find this Gaussian assumption sufficient for this work. Next, we re-assign the mean of this distribution by sampling from a multivariate Gaussian centered at the ground truth in both $\xi_{\text{lens}}$ and $\Delta \phi$, with the same covariance matrix. This step is necessary to guarantee that the emulated posteriors are well calibrated and unbiased, and removes any influence from the interim modeling prior used in the NPE method: $p(\Delta\phi, \xi_{\text{lens}}|d_{\text{img}},\nu_{\text{int}})$ $\rightarrow$ $p(\Delta\phi, \xi_{\text{lens}}|d_{\text{img}})$.

From these calibrated models, we use a simple re-scaling of the covariance matrix of $p(\Delta\phi, \xi_{\text{lens}}|d_{\text{img}})$ to emulate four different image modeling scenarios. Increasing from lowest precision to highest precision, we emulate: automated modeling of LSST images (LSST-NPE), automated modeling of HST images (HST-NPE), dedicated forward modeling of HST images (HST-FM), and dedicated forward modeling of JWST images (JWST-FM). When generating the initial NPE models, we run on both HST and LSST images, which results in a $p(\Delta\phi, \xi_{\text{lens}}|d_{\text{img}})$ for both the HST-NPE and LSST-NPE scenarios without needing to re-scale. To emulate the HST-FM models, we re-scale the HST-NPE covariance matrix such that there is an uncertainty of 0.04 on the power-law slope, roughly matching the modeling precision on the HST image models in \citetalias{TDCOSMO_2025}. To emulate the JWST-FM models, we re-scale the HST-NPE covariance matrix for a 2$\%$ constraint on the Fermat potential difference, based on the work of \cite{williams_2025}.

Finally, we combine samples of $\xi_{\text{lens}}$ from $p(\Delta\phi, \xi_{\text{lens}}|d_{\text{img}})$ with samples from an interim prior over $\beta_{\text{ani}}$, $p(\beta_{\text{ani}}|\nu_{\text{int}})$, to pre-compute the kinematic model quantity, $\mathcal{J}(\xi_{\text{lens}},\beta_{\text{ani}})$. We use the spherical Jeans anisotropy modeling (JAM) implemented in \textsc{lenstronomy} \citep{birrer2018lenstronomy}. For the anisotropy interim prior, we use a Gaussian with $\nu_{\text{int}} = \{\mu(\beta_{\text{ani}})=0,\sigma(\beta_{\text{ani}})=0.2\}$, truncated at -0.5,+0.5. This prior is motivated by measurements of early-type galaxies in \cite{cappellari2025_ETGs}. As shown in Equation \ref{eqn:lklhd_integral_full}, this informative prior assumption is fully accounted for in the hierarchical inference. But, we still use a physically motivated prior because the prior influences where the density of importance samples is highest. After computing all relevant quantities, we fit a Gaussian to 500 samples over all parameters: $p(\Delta\phi, \xi_{\text{lens}}|d_{\text{img}})$ $\rightarrow$ $p(\mathcal{J}, \beta_{\text{ani}},\Delta\phi,\xi_{\text{lens}}|d_{\text{img}},\nu_{\text{int}})$. Note the re-introduction of $\nu_{\text{int}}$ is used to track the dependence on the interim importance sampling distribution for $\beta_{\text{ani}}$. This final distribution tracks both the final emulated mass model posterior, $p(\xi_{\text{lens}}|d_{\text{img}})$, and a Gaussian fit to the relationship between those mass model parameters and the quantities of interest: $\Delta \phi(\xi_{\text{lens}})$ and $\mathcal{J}(\xi_{\text{lens}},\beta_{\text{ani}})$. This process is approximate, and assumes the relationship can be approximated by a multivariate Gaussian. This assumption allows us to quickly generate thousands of new importance samples and their associated model quantities from the original set of 500. In future applications, it may be necessary to instead directly compute the model quantities for every importance sample of $\xi_{\text{lens}}$. This would increase the computational cost for data vector generation, but would not change the final cosmological likelihood evaluation time.

After emulating all four model types, we assess the median precision on the Fermat potential difference, $\Delta \phi$, across the lens sample for each modeling fidelity. We report median precisions of 2\% for JWST-FM, 4\% for HST-FM, 11\% for HST-NPE, and 18\% for LSST-NPE.

\begin{figure*}[hbt!]
    \centering
    \includegraphics[scale=0.35]{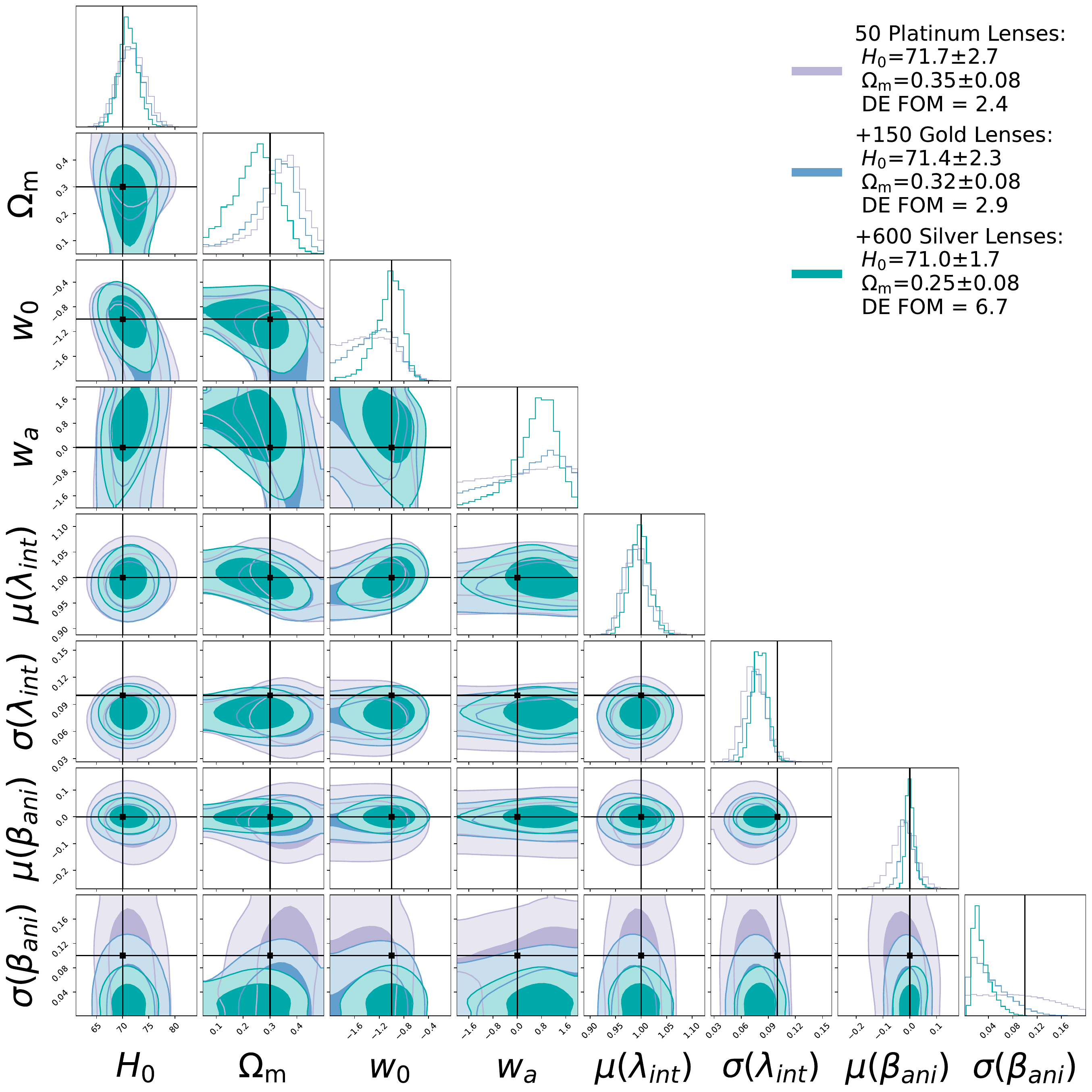}
    \caption{ We demonstrate how growing the sample size impacts the cosmological inference from the baseline experiment configuration (Table \ref{table:baseline}). We start with 50 platinum lenses only (light purple), then add in 150 gold lenses with LSST time-delay measurements (light blue), and finally add in 600 silver lenses with  LSST time-delays and image-based mass models (blue-green). We see that the addition of the 750 lenses that depend on LSST measurements simultaneously improves the DE FOM from 2.4 to 6.7, and the $H_0$ precision from $\sim$3.5\% to $\sim$2.5\%.}
    \label{fig:baseline}
\end{figure*}

\subsection{Time Delay Measurement}
\label{subsection:time_delays}

We assume a time-delay measurement with Gaussian uncertainty: $\mu_{obs}(\mathbf{\Delta t})$, $\Sigma_{obs}(\mathbf{\Delta t})$. For this work, we simply assign the size of the measurement error based on the assumed fidelity of time-delay measurement. We leave the full simulation and modeling of light-curves for future work. For lenses with long-term monitoring, we assume a $3\%$ measurement, based on measurements used in \citetalias{TDCOSMO_2025}. From LSST light-curves, we take a conservative baseline assumption of a 5-day precision measurement, typical of what has been achieved from previous low-cadence ($\sim $4 days) decade-long single-band monitoring campaigns \citep[e.g.][]{Millon2020}. However, a precision of 1 day can probably be reached for most lensed systems after 10 years, if light curves can be jointly modeled across photometric bands. This issue is left for future work and will be addressed in the next papers of our series. We note that we assign uncertainties at a fixed percentage for the lenses with long-term monitoring (3$\%$), and at a fixed absolute amount (5 days) for the larger sample. This is due to the difference in observing strategy. In the LSST survey, the same cadence is applied uniformly, resulting in the same absolute uncertainty for every lens. In dedicated campaigns, the cadence can be optimized to reach a particular percent precision per lens. 

\subsection{Stellar Kinematics}
\label{subsection:stellar_kin}

We anticipate many lenses will have spectroscopic follow-up to determine the velocity dispersion profile of the lensing galaxy. Our assumptions for the follow-up fidelity are motivated by current best practices established by \cite{knabel_2025}.

Some lenses will be observed at length with IFU. In this setting, the velocity dispersion profile can be constrained in radial bins, and the change in velocity dispersion at different radii can be traced. We assume two options for this kind of spatially-resolved kinematic follow-up. The highest fidelity is measurement by the JWST NIRSpec instrument. The second highest fidelity is by the ground-based VLT-MUSE or Keck-KCWI spectrographs. For all measurements, we assume independent Gaussian measurement errors over a vector of velocity dispersions in each bin: $\mu_{obs}(\bm{\sigma_v})$, $\Sigma_{obs}(\bm{\sigma_v})$. For lenses with resolved spectroscopic measurements coming from the NIRSpec instrument on JWST, we assume 10 radial bins of width 0.2", with bin edges going from 0" to 2". We assume a Gaussian PSF with FWHM=0.05". We match the number of bins assumed in previous forecasts that incorporate JWST kinematics \citep{tdcosmo_V}. This assumption is an approximation, as the first measurements of spatially-resolved kinematics from JWST NIRSpec are just being delivered now \citep{first_jwst_kin}. For this work, we assume lenses with second-highest fidelity are observed with VLT-MUSE, given its location in the Southern hemisphere. We assume three radial bins of width 0.5", with bin edges going from 0" to 1.5". We assume a Gaussian PSF with FWHM=0.5", which is an optimistic estimate given the range of seeings reported in previous campaigns \citep{desi_muse_obs}. In both cases, we simplify our emulation to use static bin sizing, regardless of the Einstein radius of the lens, assuming some de-blending of spectral components will be applied. More detailed treatment of spatially-resolved kinematics, including dynamic bin sizing and correlated measurement errors, will be incorporated into future work. 

A larger number of time-delay lenses will have single-aperture stellar kinematic observations. This results in a single velocity dispersion measurement, which anchors the mass profile but does not constrain radial changes. Here, we also assume a Gaussian measurement error: $\mu_{obs}(\sigma_v)$, $\sigma_{obs}(\sigma_v)$. The ChANGES survey, operated by the 4MOST fiber instrument on the VISTA telescope, has time allocated for $\sim$1500 lensed AGN discovered by Rubin \citep{CHANGES_survey}. Additional follow-up from other instruments, such as Magellan IMACS, ESO-NTT SOXS, and VLT FORS2, will likely complement the survey observations from 4MOST. For this work, all single-aperture observations are assumed to be done with 4MOST. There are key differences between the different measurement options, but we assume this effect is sub-dominant for this work. For 4MOST measurements, we use a single aperture of R = 0.725", and a Gaussian PSF with fwhm=0.5".

\subsection{Line-of-Sight Convergence}
\label{subsection:ext_conv}

Each lens also needs a measurement of the lensing environment to constrain the external convergence, $\kappa_{\text{ext}}$. We assume we have a Gaussian posterior for $\kappa_{\text{ext}}$ of width 0.05 for each lens, typical of the current precision achieved by the most recent LOS modeling techniques \citep{Rusu2017, Wells_2024}. The LOS measurement only requires deep multi-band imaging around each lens, which will be obtained by LSST.

\section{Experiments}
\label{section:experiments}

Given a simulated catalog of LSST lenses, our goal is to estimate the cosmological constraining power in the LSST sample. We start by building a baseline lens sample, based on existing campaigns and best estimations (see Section \ref{subsection:baseline_exp} and Table \ref{table:baseline}). Then, we run experiments to assess how cosmological constraining power depends on follow-up fidelity. We investigate kinematics in Section 
\ref{subsection:kin_exp}, with experiments summarized in Table \ref{table:kinematics}. We investigate image-based mass models in Section \ref{subsection:massmodel_exp}, with experiments summarized in Table \ref{table:mass_models}. We investigate time-delay measurements in Section \ref{subsection:timedelay_exp}, with experiments summarized in Table \ref{table:time_delays}. We also investigate how the constraint depends on the redshift configuration of the lenses in Section \ref{subsection:redshift_config}.

In all experiments, the final posterior is taken from 50,000 MCMC samples (see Section \ref{subsection:HI_method}). We start from a chain of 70,000 samples, and remove the first 20,000 samples for burn-in. To visualize posterior distributions from these samples, we use the \textsc{corner}\footnote{\url{https://github.com/dfm/corner.py}} package \citep{corner_py}. To compare performance across experiments, we use the Dark Energy Figure of Merit (DE FOM) \citep{DETF}. This metric assumes a Gaussian posterior over ($w_0$,$w_a$). Our DE FOM calculation is approximate, as we use a Gaussian fit to the posterior samples. First, we compute the pivot scale-factor, $a_p$, where the uncertainty on w(a) is minimized:
\begin{equation}
    1 - a_p = - \frac{\langle \delta w_0 \delta w_a \rangle}{\langle \delta w_a^2 \rangle }
\end{equation}
Then, we re-parameterize the posterior from ($w_0$,$w_a$) to ($w_p$,$w_a$), where: 
\begin{equation}
    w_p  = w_0 + (1 - a_p)w_a
\end{equation}
Under this parameterization, the 2-Dimensional posterior area does not have a tilt, and the inverse area of the 68$\%$ interval of the posterior can be computed as: 
\begin{equation}
    \text{DE FOM} = \frac{1}{\sigma(w_p)\sigma(w_a)}.
\end{equation}

This value is equivalent to the 68$\%$ area in the ($w_0$, $w_a$) parameterization as well \citep{desc_srd}. A higher DE FOM value indicates higher precision on the equation of state parameters. We report both the DE FOM and the pivot redshift, $z_p$. We choose to report dark energy precision with the DE FOM, rather than using the 1D intervals on ($w_0$,$w_a$), since this metric accounts for a changing pivot redshift. To assess precision on $H_0$ and $\Omega_{\text{m}}$, we define $\sigma(x)$ as one half of the width of the 68$\%$ highest density interval (HDI) of the 1D posterior in x. In a Gaussian posterior, this corresponds to the 1$\sigma$ value. We use this definition to account for non-symmetric posteriors.

\subsection{Baseline}
\label{subsection:baseline_exp}

For our first experiment, we build a baseline of what we expect the sample of LSST lensed AGN will look like. We assume a total sample of 800 lenses. Only a small number of these lenses will receive the highest fidelity follow-up. A larger portion of the sample will have LSST data products only. We make assumptions for what the breakdown will look like, based on existing campaigns, and our best estimates (Table \ref{table:baseline}). We investigate the impact of the assumed follow-up fidelity in further experiments. 

\begin{table}[hbt]
\centering
\begin{tabular}{| c | c | c |}
    \hline
    
    \textbf{Experiment} & \# Lenses & Kinematic Type \\
    \hline \hline

    \multicolumn{3}{|c|}{\textbf{Experiment 1.1: Extra IFU Kinematics}} \\
    \hline
    Platinum-JWST & 10 & NIRSpec \\
    \textbf{Platinum-VLT} & 40 $\rightarrow$ \textbf{112} & \textbf{MUSE}  \\
    \textbf{Gold-4MOST} & 150 $\rightarrow$ \textbf{78} & \textbf{4MOST/Magellan}  \\
    Silver-4MOST & 300 & 4MOST/Magellan \\
    Silver & 300 & None  \\
    \hline

    \multicolumn{3}{|c|}{\textbf{Experiment 1.2: Extra Aperture Kinematics}} \\
    \hline
    Platinum-JWST & 10 & NIRSpec \\
    Platinum-VLT & 40 & MUSE \\
    Gold-4MOST & 150 & 4MOST/Magellan \\
    \textbf{Silver-4MOST} & 300 $\rightarrow$ \textbf{600} & \textbf{4MOST/Magellan} \\
    \textbf{Silver} & 300 $\rightarrow$ \textbf{0} & \textbf{None} \\
    \hline

\end{tabular}
\caption{Experiments with variations in kinematic follow-up. We show how these experiments change from the baseline configuration, detailed in Table \ref{table:baseline}. Quantities not listed here stay the same as the baseline.}
\label{table:kinematics}
\end{table}

\begin{table}[hbt]
\centering
\begin{tabular}{| c | c | c |}
    \hline
    
    \textbf{Experiment} & \# Lenses & Image Model Type \\
    \hline \hline

    \multicolumn{3}{|c|}{\textbf{Experiment 2.1: Extra Space-Based Imaging}} \\
    \hline
    Platinum-JWST & 10 & JWST-FM \\
    Platinum-VLT & 40 & HST-FM \\
    Gold-4MOST & 150 & HST-NPE \\
    \textbf{Silver-4MOST} & \textbf{300} & \textbf{HST-NPE} \\
    Silver & 300 & LSST-NPE \\
    \hline

    \multicolumn{3}{|c|}{\textbf{Experiment 2.2: Extra Forward Modeling}} \\
    \hline
    Platinum-JWST & 10 & JWST-FM \\
    Platinum-VLT & 40 & HST-FM \\
    \textbf{Gold-4MOST} & \textbf{150} & \textbf{HST-FM} \\
    Silver-4MOST & 300 & LSST-NPE \\
    Silver & 300 & LSST-NPE \\
    \hline

    \multicolumn{3}{|c|}{\textbf{Experiment 2.3: Extra FM + Extra Imaging}} \\
    \hline
    Platinum-JWST & 10 & JWST-FM \\
    Platinum-VLT & 40 & HST-FM \\
    \textbf{Gold-4MOST} & \textbf{150} & \textbf{HST-FM} \\
    \textbf{Silver-4MOST} & \textbf{300} & \textbf{HST-FM} \\
    Silver & 300 & LSST-NPE \\
    \hline

\end{tabular}
\caption{Summary of experiments with variations in mass model fidelity. We show how these experiments change from the baseline configuration, detailed in Table \ref{table:baseline}. Quantities not listed here stay the same as the baseline.}
\label{table:mass_models}
\end{table}

\begin{table}[hbt]
\centering
\begin{tabular}{| c | c | c |}
    \hline
    
    \textbf{Experiment} & \# Lenses & Time-Delay Precision  \\
    \hline \hline

    \multicolumn{3}{|c|}{\textbf{Experiment 3.1: Extra Long-Term Monitoring}} \\
    \hline
    Platinum-JWST & 10 & 3\% \\
    Platinum-IFU & 40 & 3\%  \\
    \textbf{Gold-4MOST-LTM} & + \textbf{60} & \textbf{2 days}\\
    \textbf{Gold-4MOST} & 150 $\rightarrow$ \textbf{90} & \textbf{5 days}\\
    Silver-4MOST & 300 & 5 days \\
    Silver & 300 & 5 days  \\
    \hline

    \multicolumn{3}{|c|}{\textbf{Experiment 3.2: $\sigma(\Delta t)_{LSST} = $ 4 days}} \\
    \hline
    Platinum-JWST & 10 & 3\% \\
    Platinum-VLT & 40 & 3\% \\
    \textbf{Gold-4MOST} & 150 & \textbf{4 days} \\
    \textbf{Silver-4MOST} & 300 & \textbf{4 days} \\
    \textbf{Silver} & 300 & \textbf{4 days} \\
    \hline

    \multicolumn{3}{|c|}{\textbf{Experiment 3.3: $\sigma(\Delta t)_{LSST} = $ 3 days}} \\
    \hline
    Platinum-JWST & 10 & 3\% \\
    Platinum-VLT & 40 & 3\% \\
    \textbf{Gold-4MOST} & 150 & \textbf{3 days} \\
    \textbf{Silver-4MOST} & 300 & \textbf{3 days} \\
    \textbf{Silver} & 300 & \textbf{3 days} \\
    \hline

    \multicolumn{3}{|c|}{\textbf{Experiment 3.4: $\sigma(\Delta t)_{LSST} = $ 2 days}} \\
    \hline
    Platinum-JWST & 10 & 3\% \\
    Platinum-VLT & 40 & 3\% \\
    \textbf{Gold-4MOST} & 150 & \textbf{2 days} \\
    \textbf{Silver-4MOST} & 300 & \textbf{2 days} \\
    \textbf{Silver} & 300 & \textbf{2 days} \\
    \hline

\end{tabular}
\caption{Summary of experiments with variations in time-delay precision. We show how these experiments change from the baseline configuration, detailed in Table \ref{table:baseline}. Quantities not listed here stay the same as the baseline.}
\label{table:time_delays}
\end{table}

We assume 10 lenses will have the most expensive lens follow-up, which is JWST NIRSpec IFU kinematics (``Platinum-JWST" in Table \ref{table:baseline}).  These lenses will have spatially resolved kinematic maps from NIRSpec, JWST-FM mass model quality, and long-term time delay monitoring. We assume that these lenses will have properties similar to the TDCOSMO sample \citepalias{TDCOSMO_2025}. We introduce a selection criteria for these 10 lenses, randomly selecting 10 systems from our catalog that have: four point source images, at least one time-delay longer than 30 days, lens light apparent magnitude brighter than 24, and source light apparent magnitude brighter than 24. The requirement for lens light apparent magnitude is to ensure enough signal-to-noise to constrain spatially-resolved kinematics, and the requirement on source light apparent magnitude is to account for the preference to model systems with visible host galaxy arcs. 

Next, we assume 40 lenses will have spatially resolved kinematic maps from MUSE, imaging from HST, and long-term time-delay monitoring (``Platinum-VLT" in Table \ref{table:baseline}). We randomly select these 40 lenses from the catalog with the following criteria: half of the lenses have four point source images,  at least one time-delay longer than 30 days, lens light apparent magnitude brighter than 22, and source light apparent magnitude brighter than 24. Since MUSE is a ground-based experiment, we have a stricter requirement on the lens light magnitude.

We assume 150 lenses will have both 4MOST aperture kinematics and HST imaging (``Gold-4MOST" in Table \ref{table:baseline}). This portion of the sample has time-delay measurements from LSST light-curves. The only selection criteria for these lenses is that half of the lenses have four images, again accounting for the existing selection that favors quads for image-based modeling.

Next, we move to the largest portion of the sample which has LSST imaging only. For this portion of the sample, we do not apply any selection cuts, and randomly sample from the remaining lenses in the catalog. This portion of the sample is assumed to contain 600 lenses, where 300 lenses have 4MOST aperture kinematics (``Silver-4MOST" in Table \ref{table:baseline}), and 300 have no kinematic constraint (``Silver" in Table \ref{table:baseline}). All lenses in this portion of the sample have mass models from LSST-imaging, and LSST time-delay measurements.

With this baseline configuration, we assess how adding in a larger and larger number of lenses improves cosmological constraining power. We first test with only the 50 platinum lenses, then add in the 150 gold lenses, and then finally add in the 600 silver lenses. We show the full posteriors in Figure \ref{fig:baseline}. We see that the addition of the 600 lenses from the LSST survey to the sample is crucial, simultaneously improving the DE FOM from 2.9 to 6.7, and the $H_0$ precision from $\sim$3\% to $\sim$2.5\%. We discuss further takeaways in Section \ref{subsection:discuss_statistical_sample}. We note that the recovery of the scatter in the mass-sheet and anistropy populations, $\sigma(\lambda_{\text{int}})$ and $\sigma(\beta_{\text{ani}})$, is shifted towards lower values. We hypothesize this is due to the difficulty of disentangling scatter coming from the width of the population versus scatter coming from large measurement uncertainties, especially if the effective precision on $\lambda_{\text{int}}$ and $\beta_{\text{ani}}$ approaches the width of the population distribution.

\begin{figure*}[hbt!]
\begin{center}
\subfloat[Experiment 1.1: +72 VLT IFU Kinematics]{\includegraphics[scale=0.3]{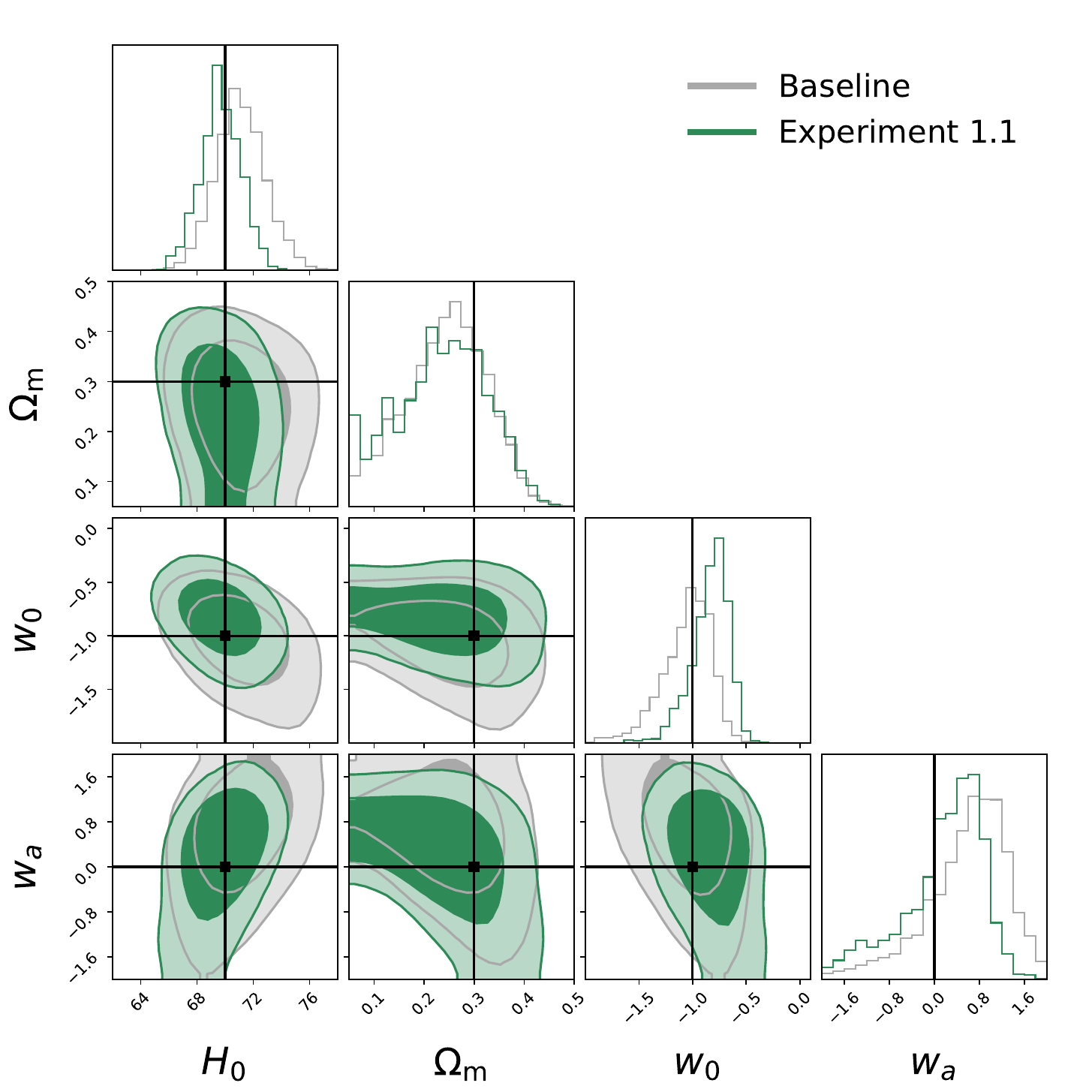}
\label{fig:exp2_1_contour}}
\quad
\subfloat[Experiment 1.2: +300 4MOST Kinematics]{\includegraphics[scale=0.3]{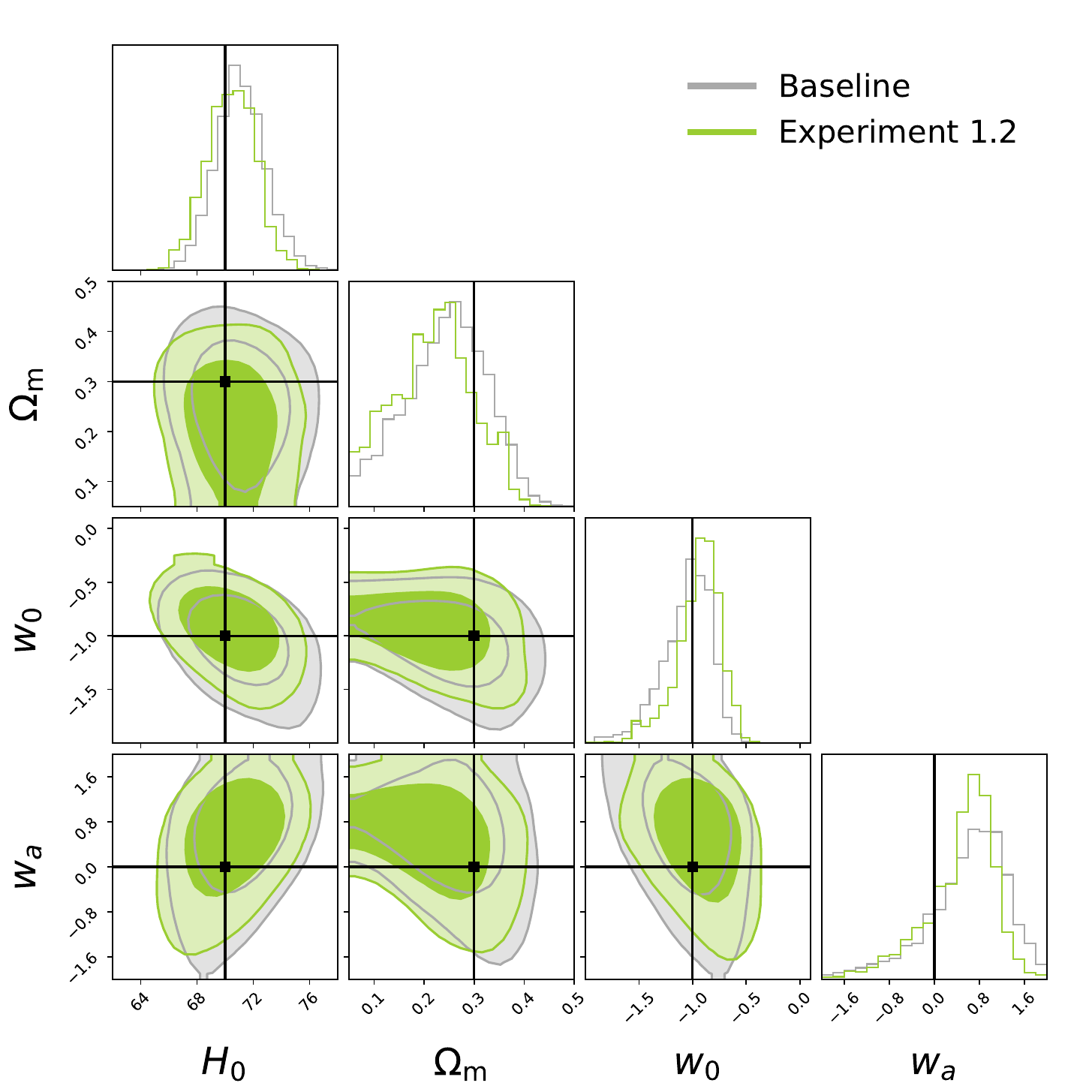}
\label{fig:exp2_2_contour}}
\quad
\subfloat[Experiment 3.1: +60 $\sigma(\Delta t) = $ 2 days]{\includegraphics[scale=0.3]{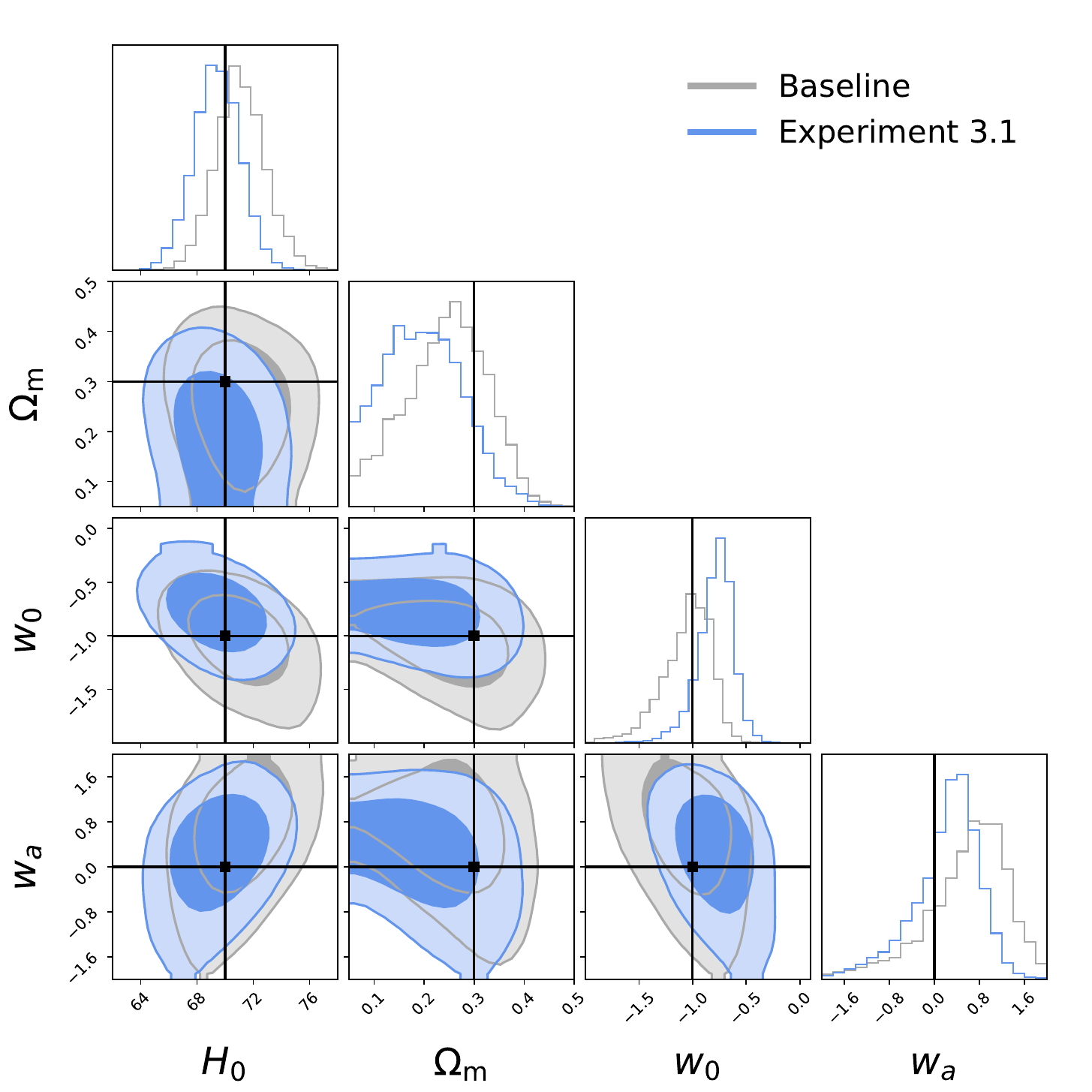}
\label{fig:exp3_1_contour}}
\quad
\subfloat[Experiment 3.4: $\sigma(\Delta t)_{LSST} = $ 2 days]{\includegraphics[scale=0.3]{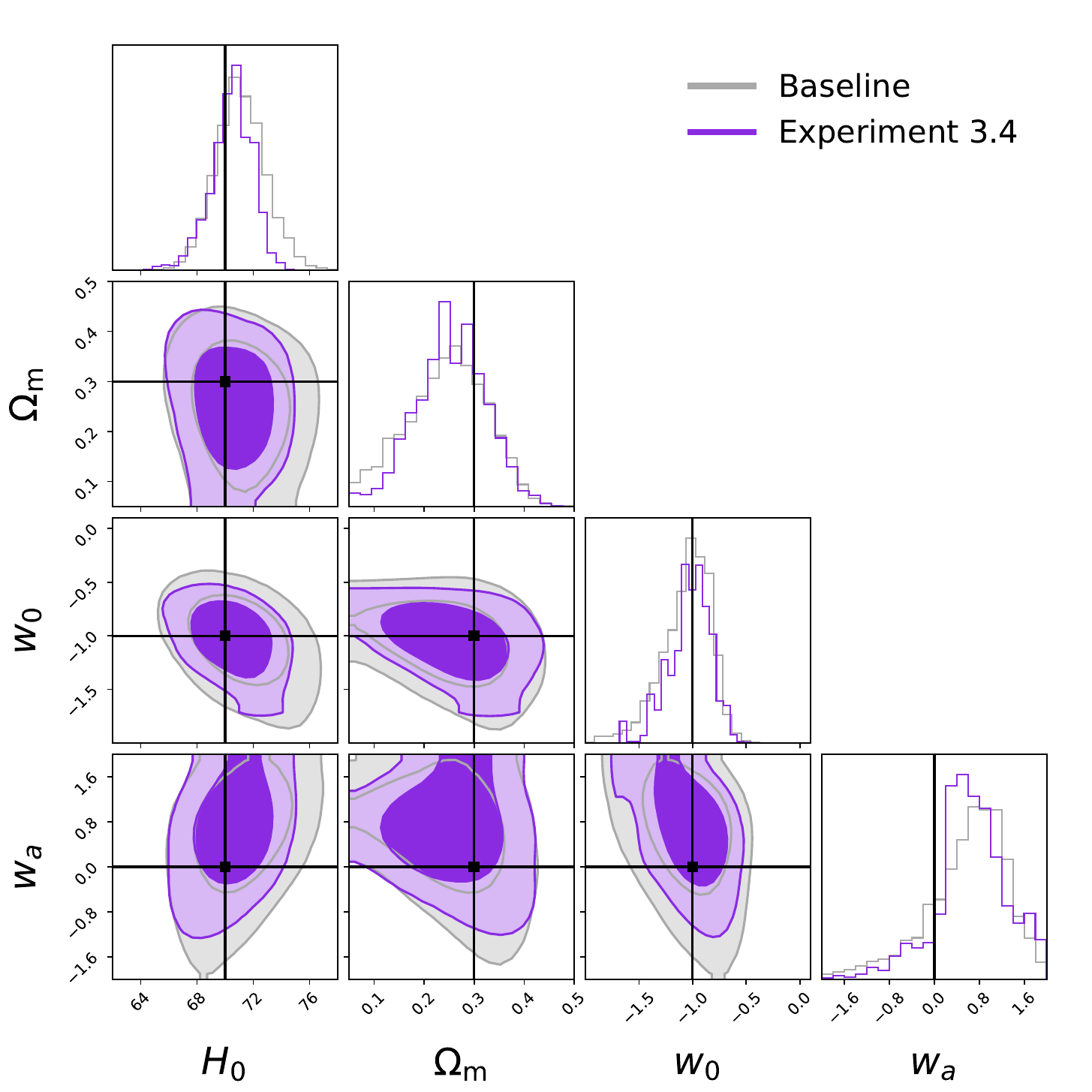}
\label{fig:exp3_4_contour}}
\caption{Cosmological posteriors from four experiments, testing additional follow-up compared to the baseline. The baseline posterior is shown in grey. On the top left, we test adding IFU observation on 72 lenses (dark green). On the top right, we test adding aperture kinematics on 300 lenses (light green). On the bottom left, we test a conservative long-term monitoring campaign, increasing the time-delay precision to 2 days for 60 lenses (light blue). On the bottom right, we test a more optimistic scenario, assuming a 2-day time-delay precision for all LSST measurements (purple).}
\label{fig:experiment_contours}
\end{center}
\end{figure*}

\begin{table*}[hbt]
\centering
\begin{tabular}{| c | c || c | c | c | c |}
    \hline
     \;\; Experiment \;\; & \;\;\;\;\;\;\;\;\;\;\;\;\;\;\;\;\;\; Description \;\;\;\;\;\;\;\;\;\;\;\;\;\;\;\;\;\; & \;\;\; $\sigma(H_0)$ \;\;\; & \;\;\; $\sigma(\Omega_{\text{m}})$ \;\;\; & \;\;\; DE FOM \;\;\; & \;\;\; $z_p$ \;\;\; \\ 
    \hline
    \hline
    \hyperref[table:baseline]{0.1} & Baseline & 1.7 & 0.08 & 6.7 & 0.17 \\ 
    \hline
    \hline
    \hyperref[table:kinematics]{1.1} & Extra IFU Kinematics & 1.4 & 0.09 & 9.1 & 0.14 \\
    \hyperref[table:kinematics]{1.2} & Extra Aperture Kinematics & 1.7 & 0.08 & 8.9 & 0.21 \\
    \hline
    \hline
    \hyperref[table:mass_models]{2.1} & Extra Space-Based Imaging & 2.0 & 0.05 & 3.5 & 0.08 \\
    \hyperref[table:mass_models]{2.2} & Extra Forward Modeling & 1.7 & 0.08 & 6.5 & 0.14 \\
    \hyperref[table:mass_models]{2.3} & Extra FM + Extra SB Imaging & 1.7 & 0.06 & 4.9 & 0.11\\
    \hline
    \hline
    \hyperref[table:time_delays]{3.1} & Extra Long-Term Monitoring & 1.6 & 0.08 & 12.0 & 0.21\\
    \hyperref[table:time_delays]{3.2} & $\sigma(\Delta t)_{LSST} = $ 4 days & 1.7 & 0.08 & 7.2 & 0.17\\
    \hyperref[table:time_delays]{3.3} & $\sigma(\Delta t)_{LSST} = $ 3 days & 1.7 & 0.09 & 7.4 & 0.14\\
    \hyperref[table:time_delays]{3.4} & $\sigma(\Delta t)_{LSST} = $ 2 days & 1.3 & 0.07 & 10.1 & 0.24\\
    \hline
\end{tabular}
\caption{Results from all experiments testing how cosmological precision improves with additional follow-up campaigns. To compute $\sigma(H_0)$ and $\sigma(\Omega_{\text{m}})$, we take one half of the 68$\%$ HDI, to account for non-symmetric posteriors. We report an approximate DE FOM = [$\sigma(w_p) \sigma(w_a)$]$^{-1}$, with pivot redshift $z_p$.}
\label{tab:metrics_all_exps}
\end{table*}

\subsection{Kinematics on a Fixed Telescope Budget}
\label{subsection:kin_exp}

We consider how additional kinematic follow-up compared to the baseline configuration can improve constraining power. To constrain dynamical deviations from a cosmological constant, we need observational data across redshifts. With this in mind, we test whether it would be better to observe a smaller number of lenses with IFU, optimizing for a better constraint of the mass-anisotropy degeneracy, or a larger number of lenses with aperture kinematics, optimizing for better redshift sampling.

With finite observing resources, one could imagine choosing between running an IFU kinematics follow-up campaign on a larger telescope, or an aperture kinematics follow-up campaign on a smaller telescope. To test this scenario, we assume we have a fixed budget of 6000 minutes on an 8-meter class telescope, which corresponds to approximately 9000 minutes on a 6.5-meter telescope to achieve the same signal-to-noise ratio, under similar observing conditions. In the first scenario, telescope time is used to upgrade Gold-4MOST lenses to become Platinum-VLT lenses, adding  spatially-resolved kinematics for 72 lenses (assuming 5000s per lens on VLT-MUSE) (Experiment 1.1 in Table \ref{table:kinematics}). The second option is to use the same amount of resources to upgrade 300 lenses from having no kinematics to having 4MOST-like aperture kinematics, assuming 30 minutes of observation on a 6.5m telescope (ex: Magellan Baade and Clay Telescopes) (Experiment 1.2 in Table \ref{table:kinematics}). 

We show the posterior over cosmological parameters with additional IFU kinematics and additional aperture kinematics in Figures \ref{fig:exp2_1_contour} and \ref{fig:exp2_2_contour}, with corresponding precision recorded in Table \ref{tab:metrics_all_exps}. We find that both strategies result in an improved constraint on DE, improving from the baseline DE FOM (6.7) to 9.1 and 8.9 respectively. For $H_0$ precision, the IFU strategy is advantageous, improving off the baseline from a $\sim$2.5\% to a $\sim$2\% constraint, whereas the aperture strategy does not improve the $H_0$ precision. We discuss the implications for kinematic follow-up strategies in Section \ref{subsection:followup_discussion}.

\subsection{Varying Mass Model Precision}
\label{subsection:massmodel_exp}

The precision of image-based mass models also plays into the error budget of TDC. We design experiments to test how improving the fidelity of the image-based models changes the cosmological constraining power. We caution that the emulation of the mass models is approximate, and results here are sensitive to the assumptions made (see Section \ref{subsection:discussion_mass_models}).

First, we assess how upgrading the imaging quality from ground-based to space-based, while keeping the modeling method automated, changes the constraining power (Experiment 2.1 in Table \ref{table:mass_models}). For the 300 lenses in the ``Silver-4MOST" category, we upgrade the mass models from LSST-NPE to HST-NPE. Next, we investigate how extra investigator effort to produce high-precision forward models (FM), using the same imaging data, could improve the constraining power (Experiment 2.2 in Table \ref{table:mass_models}). For the 150 lenses in the ``Gold-4MOST" category, we upgrade the mass models from HST-NPE to HST-FM. Note the HST-FM models are significantly more precise, with a median Fermat potential precision of 4\% compared to 11\%. But, producing this many forward models represents significant effort, as current forward modeling techniques require a few months of investigator time and up to half-million CPU hours per lens, even when automating some part of the modeling procedure (\citet{dolphin_modeling}, \citet{schmidt_STRIDES}). Finally, we assess the combination of both efforts, by upgrading both the data quality and the modeling method. We upgrade all ``Silver-4MOST" and ``Gold-4MOST" lenses to HST-FM models (Experiment 2.3 in Table \ref{table:mass_models}). Note this is an optimistic scenario, as it would require space-based imaging and dedicated FM effort for an additional 450 lenses. But, ongoing advancements in gradient-accelerated lens modeling techniques may soon enable high-precision modeling at this scale \citep[e.g.][]{Gu_2022, Galan_2022, lombardi_gravity_jl}.

We tabulate the cosmological precision for all three experiments, and compare them to the baseline, in Table \ref{tab:metrics_all_exps}. In experiments 2.1 and 2.3, which upgrade the mass modeling precision on 300 and 450 lenses respectively, we find improved $\sigma(\Omega_{\text{m}})$ (0.05 and 0.06) compared to the baseline (0.08) accompanied by a lower DE FOM (3.5 and 4.9) compared to the baseline (6.7). This result underscores the importance of assessing precision across all parameters simultaneously. Unlike other experiments, when we upgrade lenses from LSST-NPE to HST-NPE models, we effectively introduce a different statistical draw for the measurement errors of each lens, by using a different run of the NPE modeling framework. This impacts the preferred cosmological model, and thus the relative constraining power between $\Omega_{\text{m}}$ and DE, as well as the preferred central value, which also impacts the DE FOM, as discussed in Appendix \ref{appendix:random_seeds}. A more controlled experiment is Experiment 2.2, where the same measurements are used (HST-NPE), but are re-scaled to higher precision (HST-FM). In this experiment, assessing our metrics in Table \ref{tab:metrics_all_exps}, we see no significant change compared to the baseline in any parameter. This indicates that at the time-delay and kinematic precision assumed, the mass model precision is not a limiting factor. We discuss further in Section \ref{subsection:followup_discussion}.

\subsection{Varying Time-Delay Precision}
\label{subsection:timedelay_exp}

We are also interested in how the time-delay measurement precision impacts the cosmological constraining power. First, we investigate how extra long-term monitoring can be used to improve the DE FOM. Then, we assess how changing the precision for time-delay measurements from LSST light curves impacts the DE FOM. 

To investigate the impact of additional long-term monitoring, we assume 60 of the gold lenses with 4MOST aperture kinematics will receive additional long-term monitoring, reducing the time delay measurement uncertainty on those lenses to 2 days (Experiment 3.1 in Table \ref{table:time_delays}). We assume a 2 day precision to emulate a follow-up campaign similar to the one performed by \citet{tdcosmo_xvii}. Next, to understand how the measurement precision from LSST light curves affects constraining power, we improve the measurement error for all lenses with LSST time-delays. We improve precision from the baseline of 5 days to 4 days (Experiment 3.2), 3 days (Experiment 3.3), and 2 days (Experiment 3.4, see Table \ref{table:time_delays}).

We tabulate results from these experiments in Table \ref{tab:metrics_all_exps}. In experiment 3.1, where we increase the time-delay precision to 2 days on 60 lenses, the DE FOM improves from the baseline of 6.7 to 12.0, $\sigma(H_0)$ improves slightly from the baseline of 1.7 to 1.6, and $\sigma(\Omega_{\text{m}})$ does not change from the baseline of 0.08. We show the posterior for this experiment in Figure \ref{fig:exp3_1_contour}. In Experiment 3.4, where \textit{more} lenses are upgraded to have a 2 day time-delay measurement precision, the DE FOM improves to a slightly lower value (10.1), accompanied by a significantly higher precision on both $H_0$ and $\Omega_{\text{m}}$ ($\sigma(H_0)$=1.3 and $\sigma(\Omega_{\text{m}})$=0.07). This again demonstrates the the importance of assessing constraining power across all parameters simultaneously. We also show the contour for Experiment 3.4 in Figure \ref{fig:exp3_4_contour}. When we assess how changing the time-delay precision across the LSST sample affects the inference, by comparing Experiments 3.2 and 3.3 to Experiment 3.4, we only see a significant improvement in the cosmological precision once measurement precision reaches 2 days. For 4-day and 3-day measurement precision, $H_0$ recovery does not change from the baseline of $\sigma(H_0)$=1.7, and only improves to $\sigma(H_0)$=1.3 once the measurement precision improves to 2 days. Similarly for the DE FOM, significant improvement off the baseline is only achieved with 2-day measurement precision. We discuss these results further in Section \ref{subsection:followup_discussion}.

\subsection{Investigating Redshift Configuration}
\label{subsection:redshift_config}

We design a test to investigate the impact of redshift configuration on cosmological constraining power. The strategy is to fix every property of the lens sample, except for the lens and source redshifts. We select a single lensed AGN near the median of the sample in Einstein radius, lens apparent magnitude, AGN apparent magnitude, and host galaxy apparent magnitude. We replicate that lens ten times to create a ``sample".  We assume that lens has a $1\%$ measurement on the time-delay, $2\%$ measurement on the Fermat potential, and a $1\%$ measurement on the velocity dispersion, such that the $D_{\Delta t}$ precision is constant across redshift configuration. Then, we change the redshift configuration of the sample, assigning each system a lens and source redshift drawn from narrow Gaussian populations, where $\sigma(z_{\text{lens}}),\sigma(z_{\text{src}})=0.1$. We test $\mu(z_{\text{lens}})$= 0.2, 0.5, 1 and $\mu(z_{\text{src}})$= 1, 2, 3.  These lens and source redshifts are within the ranges expected for the LSST sample, based on existing population simulations \citep{om10, abe_2025, slsim_khadka}. After assigning redshifts, we produce data vectors, and run the full hierarchical inference. We use an informative prior on the $\lambda_{\text{int}}$ and $\beta_{\text{ani}}$ population during this hierarchical inference, to emulate the inference within a larger sample. We compare the resulting cosmological precision for different redshift configurations in Table \ref{tab:redshift_exp}. First, we note that the constraining power only appears to be impacted by the lens redshift, whereas source redshift has no significant effect. With a uniform prior on $\Omega_{\text{m}}$, the DE FOM is highest at low lens redshift, with DE FOM $\sim$3 at $\mu(z_{\text{lens}})$= 0.2 and DE FOM $\sim$2 at $\mu(z_{\text{lens}})$= 1. In contrast, precision on $\Omega_{\text{m}}$ is optimized at high lens redshift, with $\sigma(\Omega_{\text{m}})$= 0.14 at $\mu(z_{\text{lens}})$= 0.2 and $\sigma(\Omega_{\text{m}})$= 0.08 at $\mu(z_{\text{lens}})$= 1. Given this behavior, we also run the experiment with an informative prior on $\Omega_{\text{m}}$, using a Gaussian approximation of the Pantheon+ prior used in \citetalias{TDCOSMO_2025}: $\mathcal{N}(\Omega_{\text{m}} | \mu=0.3,\sigma=0.018)$. With an informative $\Omega_{\text{m}}$ prior, we still find that lower lens redshift produces a higher DE FOM, with DE FOM $\sim$6 at $\mu(z_{\text{lens}})$= 0.2 and DE FOM $\sim$3.5 at $\mu(z_{\text{lens}})$= 1. We provide further discussion in Section \ref{subsection:redshift_discussion}, and show contours from the experiment in Appendix \ref{appendix:redshift}.

\begin{table*}[hbt]
\centering
\begin{tabular}{| c | c || c | c | c || c | c |}
    \hline
     \multicolumn{2}{|c||}{} & \multicolumn{3}{c||}{Uniform $\Omega_{\text{m}}$ Prior} & \multicolumn{2}{c|}{Informative $\Omega_{\text{m}}$ Prior} \\
    \hline
    \hline
      \;\;\; $\mu(z_{\text{lens}})$ \;\;\; & \;\;\; $\mu(z_{\text{src}})$ \;\;\; & \;\;\; $\sigma(H_0)$ \;\;\; & \;\;\; $\sigma(\Omega_{\text{m}})$ \;\;\; & \;\;\; DE FOM \;\;\; & \;\;\; $\sigma(H_0)$ \;\;\; & \;\;\; DE FOM \;\;\; \\ 
    \hline
    \hline
    0.2 & 1. & 0.9 & 0.14 & 2.9 & 0.9 & 6.0 \\ 
    \hline
    0.2 & 2. & 0.9 & 0.14 & 3.1 & 0.9 & 5.8\\ 
    \hline
    0.2 & 3. & 0.9 & 0.14 & 2.9 & 0.9 & 6.0 \\
    \hline
    \hline
    0.5 & 1. & 3.7 & 0.11 & 2.2 & 3.8 & 2.8 \\
    \hline
    0.5 & 2. & 3.7 & 0.11 & 2.3 & 4.1 & 2.6 \\
    \hline
    0.5 & 3. & 3.6 & 0.11 & 2.4 & 3.9 & 2.7 \\
    \hline
    \hline
    1. & 2. & 9.0 & 0.08 & 1.9 & 6.2 & 2.3 \\
    \hline
    1. & 3. & 9.2 & 0.08 & 2.0 & 6.2 & 2.4 \\
    \hline
\end{tabular}
\caption{Assessing how the redshift configuration of time-delay lenses impacts constraining power, with and without an informative $\Omega_{\text{m}}$ prior. Lens and source redshifts are drawn from a Gaussian distribution, changing $\mu(z_{\text{lens}})$ and $\mu(z_{\text{src}})$, and keeping constant $\sigma(z_{\text{lens}}),\sigma(z_{\text{src}})=$ 0.1. To compute $\sigma(H_0)$ and $\sigma(\Omega_{\text{m}})$, we take one half of the 68$\%$ HDI, to account for non-symmetric posteriors. We report an approximate DE FOM = [$\sigma(w_p) \sigma(w_a)$]$^{-1}$. We demonstrate how the lens redshift significantly impacts the cosmological constraint, while the source redshift has no significant effect.}
\label{tab:redshift_exp}
\end{table*}

\section{Discussion}
\label{section:discussion}

We emulate the full joint hierarchical inference from a LSST lensed AGN sample in multiple scenarios, investigating the potential for TDC as a DE probe. We demonstrate the improvement in constraining power with a larger sample size (Section \ref{subsection:discuss_statistical_sample}), the impact of additional follow-up campaigns (Section \ref{subsection:followup_discussion}), and the effect of redshift configuration (Section \ref{subsection:redshift_discussion}). We also compare our results to existing forecasts (Section \ref{subsection:comp_to_prev_work}), discuss limitations (Section \ref{subsection:discussion_mass_models}), and suggest future directions (Section \ref{subsection:future_work}). 

\subsection{Information in the LSST Sample}
\label{subsection:discuss_statistical_sample}

We investigate how a larger sample of lenses with lower precision measurements can work in tandem with a smaller sample of high-precision lenses, to produce the best DE constraint possible. As shown in Figure \ref{fig:baseline}, the addition of 600 lenses from the LSST sample, none of which have IFU kinematics or forward-modeling precision mass models, results in a $\sim$130\% improvement in the DE FOM (2.9 to 6.7) when compared to a 200 lens sample, and a $\sim$180\% improvement in the DE FOM (2.4 to 6.7) when compared to a 50 lens sample. We demonstrate that the gain in statistical power from incorporating many lenses with lower precision models is quite valuable, motivating further efforts to produce well-calibrated and un-biased posteriors from LSST-quality imaging.

\subsection{Follow-Up Campaigns}
\label{subsection:followup_discussion}

Our experiments are designed to inform strategies for additional observational campaigns, investigating upgrades to stellar kinematics measurements, image-based mass models, and time-delay measurements.

First, we investigate additional spectroscopic campaigns for stellar velocity dispersion measurements. In \cite{tdcosmo_V}, when testing in $\Lambda$CDM, stellar kinematics follow-up was established as the primary driver for improved constraining power on $H_0$. This is because the largest portion of the error budget in $\Lambda$CDM TDC comes from the mass-sheet and mass-anisotropy degeneracies \citep{tdc_review_24}. Our experiments in $w_0w_a$CDM agree that upgraded stellar kinematics measurements improve constraining power significantly. When expanding off a baseline lens sample that already includes 50 lenses with IFU kinematics, we are interested in knowing whether upgrading a smaller number of lenses with IFU kinematics, or a larger number of lenses with single-aperture kinematics, would be most advantageous for a DE constraint. We find that both strategies seem to be equally valuable when optimizing for the DE FOM, producing a 36$\%$ and 33$\%$ improvement off the baseline value, respectively. When additionally optimizing for the constraint on $H_0$, the IFU strategy is advantageous, improving off the baseline from a $\sim$2.5\% constraint to a $\sim$2\% constraint.

When testing improvements to the mass model precision, we do not demonstrate significant effects on cosmological constraining power, with the caveat that mass model emulation presents unique challenges compared to emulation of time-delay and velocity dispersion measurements. When comparing Experiments 2.1 and 2.3 to the baseline configuration, we see improved precision on $\Omega_{\text{m}}$, but reduced precision on DE. We hypothesize this is impacted by the effective modification of the statistical draw of measurements when switching from LSST-NPE to HST-NPE models, ultimately changing the preferred cosmological model, and thus the distribution of constraining power between $\Omega_{\text{m}}$ and DE. When upgrading from automated modeling quality to higher precision forward modeling quality in Experiment 2.2, controlling for the statistical draw of the measurements, there is no significant change to the cosmological constraint. This indicates that, given the assumed time-delay and velocity dispersion measurement precisions, the mass model precision plays a sub-dominant role in the error budget. However, as discussed in Section \ref{subsection:discussion_mass_models}, our emulation of forward model posteriors lacks full realism, and may undersell the potential improvement from forward models over automated models. Ultimately, if optimizing for DE constraining power, it is unclear from this work whether improving the mass model precision will result in significant improvement.

We find strong motivation for additional effort to improve time-delay measurements. We demonstrate that both a smaller, dedicated long-term monitoring campaign (Experiment 3.1) and an overall improvement in LSST time-delay measurement precision to 2-days (Experiment 3.4) significantly improve the cosmological constraining power compared to the baseline experiment. When we assess improving the time-delay precision across the whole LSST sample, testing 4-day, 3-day, and 2-day precision, we only see a significant improvement in the constraining power when pushing to 2-day precision, motivating further efforts to improve multi-band light-curve modeling to achieve this threshold. We also further discuss the difference in DE FOM between Experiment 3.1 and Experiment 3.4, where we improve the measurement precision to 2-days on 60 and 750 lenses respectively. While we see a lower DE FOM in Experiment 3.4 compared to Experiment 3.1, we do see a higher precision on $H_0$ and $\Omega_{\text{m}}$, underscoring the importance of assessing precision across the parameter space. An additional factor is that the central value of the $w_0$ posterior changes between the two experiments. As demonstrated in Appendix \ref{appendix:random_seeds}, the central value of the posterior also affects the resulting DE FOM, due to prior volume effects.

\subsection{Redshift Configuration}
\label{subsection:redshift_discussion}
We note that the lens redshift has a significant impact on cosmological constraining power, with lower lens redshifts producing a higher DE FOM and a smaller $\sigma(H_0)$ when assessing performance at a single redshift configuration. This trend persists with both a uniform and informative $\Omega_{\text{m}}$ prior. While performance based on these metrics favors lower lens redshift, a large sample of lenses across redshifts may still be important for constraining time-evolution properties of dark energy, especially if expanding to other parameterizations of the equation of state. Additionally, when investigating the contours shown in Appendix \ref{appendix:redshift} (Figure \ref{fig:redshift_contours}), we see that changing the lens redshift also influences the orientation of the $H_0$-$w_0$ and $w_0$-$w_a$ contours, which may be of interest when considering combined probes. But, if the goal is to constrain $H_0$, and to some extent $w_0$, we clearly demonstrate lower lens redshift to be advantageous, regardless of source redshift. We provide further discussion in Appendix \ref{appendix:redshift}.

\subsection{Comparison to Previous Work}
\label{subsection:comp_to_prev_work}

In previous work, the expected DE constraint from LSST TDC was forecasted from a sample of 236 lensed AGN \citep{shajib_SLSC}. With the caveat that the two analyses make significantly different assumptions, we compare our results in Appendix \ref{appendix:comp_to_prev}. Our work reinforces the general finding that there is much more constraining power available in time-delay lenses than what is shown in the DESC Science Requirements Document \citep{desc_srd}. Through our experiments, we build a more complete view of the potential for TDC as a DE probe.

\subsection{Limitations}
\label{subsection:discussion_mass_models}

Our work does not fully incorporate all of the complexities that will impact future analysis. 

First, we note that the emulation of mass model posteriors is limited. Unlike other portions of the analysis, the mass model inference cannot be approximated by sampling independent, Gaussian posteriors from the ground truth values, due to strong correlations between parameters. We apply NPE modeling to simulated images in order to predict a full Gaussian covariance matrix in order to account for these correlations. As shown in \citet{Erickson_2025}, well-behaved NPE can produce models that reconstruct image positions. However, it should be noted that samples from the mass model posteriors in this work are not guaranteed to produce lensing configurations where the lens model and source position reproduce the image positions exactly. In future work, we plan to improve the realism of the mass models, especially those at forward modeling precision, to take advantage of ray-tracing penalty terms and narrower, non-Gaussian posterior shapes. 

We acknowledge that the population model for the mass-sheet parameter, $\lambda_{\text{int}}$, lacks full expressivity. We do not include a term accounting for a radial dependence of $\lambda_{\text{int}}$, which will be upgraded in future work.

\subsection{Future Work}
\label{subsection:future_work}

The method developed in this work lays the foundation for the analysis pipeline for TDC within DESC. We envisage the framework here as a starting point from which the LSST-scale analysis will grow. As scalable modeling methods improve, we will start plugging in actual modeling pipelines rather than emulated ones, to keep building up the full analysis.

The vectorized likelihood evaluation implementation is a starting point, and there are many improvements that are possible. As discussed in Section \ref{subsection:likelihood_eval}, assuming every component of the likelihood is Gaussian, there is an analytical solution to the integration that will speed up the evaluation. This is already derived, and will be implemented in future work as a faster option. The implementation of the analytical solution to the integration will also serve as an important tool for cross-checking the more flexible importance sampling version of the integration. In addition to cross-checking internally, this tool can also be used as a cross-check to the \textsc{hierArc} sampling method. Since the likelihood code is currently implemented with \texttt{numpy} array operations, it can be re-written in \texttt{jax}. This will allow for even faster likelihood evaluations, and the option to use gradient informed samplers. We also plan to further improve our posterior sampling by integrating with the DESC cosmological environment. We plan to explore the application of nested sampling and Hamiltonian Monte Carlo (HMC) for our cosmological inference, as these techniques have been shown to more robustly explore complex posterior spaces \citep{albert_25, staicova_25}.

Our baseline assumptions, detailed in Table \ref{table:baseline}, are our current best estimates. As noted above, as our modeling pipelines start to take shape, we can update our assumptions, and plug in increasing levels of realism, until we are able to reach a full ``round-trip" scenario, where we start from a simulated lens population, and go through the analysis pipeline all the way back to the underlying cosmology. Having this framework running on simulations before the time-delay sample is complete is crucial to allow us to probe systematics. In particular, as we start incorporating the silver sample of lenses, we will expand the hierarchical model to incorporate informative modeling priors on quantities such as the power-law slope, explicitly accounting for distribution shifts.

\section{Conclusion}
\label{section:conclusion}

In this work, we assess how the LSST sample of lensed AGN will contribute as a DE probe. We use a series of experiments to investigate how follow-up campaigns and redshift configuration affect cosmological constraining power. To enable this work, we develop a new TDC inference framework, optimized for analysis at scale. 

We re-visit the questions we posed in the introduction.
\begin{itemize}

    \item How much cosmological constraining power is contained within the larger sample of LSST lenses when combined with the smaller, more extensively studied time-delay lens sample? 

    Answer: We find that the addition of hundreds of LSST lenses to the existing time-delay lens sample does improve the dark energy constraint significantly, from DE FOM = 2.4 to 6.7 in our baseline experiment. We demonstrate the value of incorporating the larger sample despite much lower precision per lens.

    \item How does the DE FOM depend on the portion of the lens sample with stellar kinematics? Should we use telescope time to measure stellar kinematics for many lenses in a single aperture, or fewer lenses with spatially resolved kinematics, using IFU?

    Answer: We find both strategies to be effective for improving the DE FOM, with $\sim$ 30\% improvement relative to the baseline in both cases.

    \item How sensitive is the DE FOM to mass models from high-resolution imaging versus less precise mass models, obtained either from ground-based data and/or automated modeling? What about the impact of long-term time-delay monitoring, versus time-delay measurement from LSST light curves?

    Answer: We require further experimentation, with more realistic mass models, to determine the impact of mass model precision on the DE FOM. We find that improving the time-delay measurement precision can make a significant impact. In our experiments, we note a significant gain in constraining power when the time-delay measurement precision reaches 2-days.

    \item Which redshift configurations are most advantageous for measuring DE from time-delay lenses?    

    Answer: We demonstrate that the lens redshift is the dominant factor compared to the source redshift, with lower lens redshift providing higher precision on $H_0$, and, to some extent, $w_0$, when assessing the constraint from a single redshift configuration. We also demonstrate how changing the redshift of the deflector population changes the orientations of the $H_0$-$w_0$, $w_0$-$w_a$ contours.

\end{itemize}

With a scalable inference pipeline in place, we demonstrate the DE constraint from a large sample of LSST lensed AGN. We highlight the potential increase in cosmological constraining power with more spectroscopic campaigns for stellar kinematics, and increased efforts to push to a 2-day precision on all time-delay measurements. We demonstrate the future for TDC within the DESC as a dark energy probe, building towards an analysis that will deliver the first cosmological constraints from LSST Lensed AGN.

\section{Acknowledgements}
This paper has undergone internal review in the LSST Dark Energy Science Collaboration. The internal reviewer was Timo Anguita.

SE led all aspects of the analysis. MM contributed to all aspects of the analysis. PV contributed the lens catalog and LSST mass models, additionally providing feedback on all aspects. TL and PH provided feedback on inference design, mass modeling, posterior sampling, and contributed to manuscript editing. PM provided feedback on all aspects. AJS provided feedback on analysis choices, redshift configuration test, and manuscript editing. SB provided feedback on hierarchical inference, likelihood evaluation, and manuscript editing. XYH provided feedback on hierarchical inference, kinematics treatment, and manuscript editing. TA provided extensive feedback on key results and manuscript editing. SD, NK, KN provided extensive feedback on manuscript. AR provided feedback on project goals and key results.

SE acknowledges funding from NSF GRFP-2021313357, the Stanford Data Science Scholars Program. SE and others performed work under DOE Contract DE-AC02-76SF00515. MM acknowledges support by the SNSF (Swiss National Science Foundation) through mobility grant P500PT\_203114, return CH grant P5R5PT\_225598 and Ambizione grant PZ00P2\_223738. AJS acknowledges support from NASA through the STScI grant JWST-GO-2974. SB is supported by DoE Grant DE-SC0026113. TA acknowledges support from ANID-FONDECYT Regular Project 1240105 and the ANID BASAL project FB210003. 

The DESC acknowledges ongoing support from the Institut National de 
Physique Nucl\'eaire et de Physique des Particules in France; the 
Science \& Technology Facilities Council in the United Kingdom; and the
Department of Energy and the LSST Discovery Alliance
in the United States.  DESC uses resources of the IN2P3 
Computing Center (CC-IN2P3--Lyon/Villeurbanne - France) funded by the 
Centre National de la Recherche Scientifique; the National Energy 
Research Scientific Computing Center, a DOE Office of Science User 
Facility supported by the Office of Science of the U.S.\ Department of
Energy under Contract No.\ DE-AC02-05CH11231; STFC DiRAC HPC Facilities, 
funded by UK BEIS National E-infrastructure capital grants; and the UK 
particle physics grid, supported by the GridPP Collaboration.  This 
work was performed in part under DOE Contract DE-AC02-76SF00515.

Source code for this work is publicly available in the repository \href{https://github.com/smericks/fasttdc}{\textsc{fasttdc}}. This repository
makes use of public software packages: \href{https://github.com/lenstronomy/lenstronomy}{\textsc{lenstronomy}} \citep{birrer2018lenstronomy,birrer2021lenstronomy}, \href{https://github.com/swagnercarena/paltas}{\textsc{paltas}} \citep{paltas_paper}, and \href{https://github.com/dfm/emcee}{\textsc{emcee}} \citep{emcee}.

\bibliography{main}

\appendix

\section{Hierarchical Inference Derivation}
\label{appendix:inference_derivation}

We start from the framework laid out in Section \ref{subsection:HI}, with the goal of deriving a posterior evaluation, $p(\Omega,\nu|\mathcal{D})$, that will be used in MCMC sampling. We repeat the setup of key parameters for clarity. We work within a flat $w_0w_a$CDM cosmology. We jointly infer a cosmological model,
\begin{equation}
    \Omega = \{H_0,\Omega_{\text{m}},w_0,w_a\},
\end{equation}
and a set of population-level nuisance parameters,
\begin{equation}
    \nu = \{\mu(\lambda_{\text{int}}),\sigma(\lambda_{\text{int}}),\mu(\beta_{\text{ani}}),\sigma(\beta_{\text{ani}})\}.
\end{equation}
The inference is informed by a sample of lenses, where each lens has a dataset:
\begin{equation}
    \mathcal{D}_k = \{d_{\text{img}}, d_{\text{td}}, d_{\text{kin}}, d_{\text{los}}\}.
\end{equation}

See Section \ref{subsection:HI} for further definitions of the population parameters and data products. To constrain the population model from the data products, we start with Bayes' proportionality relation:

\begin{equation}
    p(\Omega,\nu|\mathcal{D}) \propto p(\mathcal{D}|\Omega,\nu)p(\Omega,\nu).
\end{equation}

We assume that each lens provides an independent constraint: 
\begin{equation}
    p(\Omega,\nu|\mathcal{D}) \propto p(\Omega,\nu) \prod_k p(\mathcal{D}_k|\Omega,\nu).
\label{eqn:a5}
\end{equation}

From here, we expand the likelihood evaluation for a single lens, $p(\mathcal{D}_k|\Omega,\nu)$. First, we introduce a key supporting derivation in Section \ref{subsection:supporting_deriv}. Then, we expand the likelihood derivation in Section \ref{appendix:likelihood_deriv}.

\subsection{Supporting Derivation: Posterior with Interim Prior $\nu_{\text{int}}$}
\label{subsection:supporting_deriv}

Assume we have a posterior over some target parameters, $\xi_k$, given some data $d_k$, that is influenced by an interim prior over the target parameters, $p(\xi_k | \nu_{\text{int}})$. The interim prior is governed by  hyperparameters $\nu_{\text{int}}$. This posterior is written as: $p(\xi_k | d_k , \nu_{\text{int}})$. We derive the relationship between $p(\xi_k | d_k , \nu_{\text{int}})$ and the probability of the data given the target parameters: $p(d_k | \xi_k)$. We start by expanding the conditional probability in the posterior: 
\begin{equation}
    p(\xi_k | d_k , \nu_{\text{int}}) = \frac{p(\xi_k, d_k, \nu_{\text{int}})}{p(d_k, \nu_{\text{int}})}. 
\end{equation}

Then, we re-apply the conditional probability rule in the numerator and denominator:
\begin{equation}
    p(\xi_k | d_k , \nu_{\text{int}}) = \frac{p(d_k | \xi_k, \nu_{\text{int}}) p(\xi_k, \nu_{\text{int}})}{p(d_k | \nu_{\text{int}}) p(\nu_{\text{int}})}. 
\end{equation}

We expand using conditional probability in the numerator again:
\begin{equation}
    p(\xi_k | d_k , \nu_{\text{int}}) = \frac{p(d_k | \xi_k, \nu_{\text{int}}) p(\xi_k | \nu_{\text{int}}) p(\nu_{\text{int}})}{p(d_k | \nu_{\text{int}}) p(\nu_{\text{int}})}, 
\end{equation}

and cancel the $p(\nu_{\text{int}})$ terms: 
\begin{equation}
    p(\xi_k | d_k , \nu_{\text{int}}) = \frac{p(d_k | \xi_k, \nu_{\text{int}}) p(\xi_k | \nu_{\text{int}})}{p(d_k | \nu_{\text{int}})}. 
\end{equation}

Note that when $\xi_k$ is specified, $d_k$ is independent of the choice of $\nu_{\text{int}}$, so we modify the left term of the numerator: 
\begin{equation}
    p(\xi_k | d_k , \nu_{\text{int}}) = \frac{p(d_k | \xi_k) p(\xi_k | \nu_{\text{int}})}{p(d_k | \nu_{\text{int}})}. 
\end{equation}

Then, we re-arrange terms: 
\begin{equation}
    p(d_k | \xi_k) = \frac{p(\xi_k | d_k , \nu_{\text{int}}) p(d_k | \nu_{\text{int}})}{p(\xi_k | \nu_{\text{int}})}. 
\end{equation}

Finally, note that with constant $d_k$ and $\nu_{\text{int}}$, we can reduce to a proportionality relation: 
\begin{equation}
    p(d_k | \xi_k) \propto \frac{p(\xi_k | d_k , \nu_{\text{int}})}{p(\xi_k | \nu_{\text{int}})}. 
    \label{eqn:lklhd_to_posterior_prop}
\end{equation}

Ultimately, we see that the the probability of the data given the target parameters is proportional to the posterior divided by the interim prior.

\subsection{Likelihood Derivation}
\label{appendix:likelihood_deriv}

Now, we return to the original task of expanding Equation \ref{eqn:a5}. We derive the likelihood evaluation for an individual lens, k. First, we expand by assuming independence between data products:
\begin{equation}
    p(\mathcal{D}_k|\Omega,\nu) = p(d_{\text{td}} |\Omega,\nu ) p(d_{\text{img}} |\Omega,\nu ) p(d_{\text{kin}} |\Omega,\nu ) p(d_{\text{los}} |\Omega,\nu ).
\end{equation}
We cannot directly evaluate the probability of an individual lens' observables given the population-level parameters. We need to connect the data products to the population model through individual lens properties, including the lens model parameters, $\xi_{\text{lens}}$, the internal mass sheet parameter, $\lambda_{\text{int}}$, the external convergence, $\kappa_{\text{ext}}$, the orbital anisotropy parameter, $\beta_{\text{ani}}$, the lens plane image positions, $\theta_{\text{im}}$, and the redshifts, ($z_{\text{lens}}$,  $z_{\text{src}}$). The lens model parameters, $\xi_{\text{lens}}$, include lens mass parameters, lens light parameters, and the point source position in the source plane. We marginalize over all of these individual properties:
\begin{multline}
    p(\mathcal{D}_k|\Omega,\nu) = \int p(d_{\text{td}} | \Omega, \lambda_{\text{int}}, \kappa_{\text{ext}}, \Delta\phi(\xi_{\text{lens}}, \theta_{\text{im}}), z_{\text{lens}}, z_{\text{src}}) \times p(d_{\text{img}} | \xi_{\text{lens}}, \theta_{\text{im}})
    \\p(d_{\text{kin}} | \Omega, \lambda_{\text{int}},\kappa_{\text{ext}}, \mathcal{J}(\xi_{\text{lens}}, \beta_{\text{ani}},), z_{\text{lens}}, z_{\text{src}}) \times p(d_{\text{los}} |\kappa_{\text{ext}},z_{\text{lens}}, z_{\text{src}}) \times p(d_{\text{spec-z}} |z_{\text{lens}}, z_{\text{src}}) 
    \\p(\lambda_{\text{int}}, \beta_{\text{ani}} | \nu) \times p(\kappa_{\text{ext}}, \xi_{\text{lens}}, \theta_{\text{im}}, z_{\text{lens}}, z_{\text{src}}) \; \; \mathrm{d}\lambda_{\text{int}}\mathrm{d}\kappa_{\text{ext}}\mathrm{d}\beta_{\text{ani}}\mathrm{d}\xi_{\text{lens}}\mathrm{d}\theta_{\text{im}}\mathrm{d}z_{\text{lens}}\mathrm{d}z_{\text{src}}.
\label{lklhd_with_redshift}
\end{multline}
Note that the time-delay term depends on the Fermat potential, $\Delta\phi(\xi_{\text{lens}}, \theta_{\text{im}})$, and the kinematic term depends on the Jeans model quantity, $\mathcal{J}(\xi_{\text{lens}},\beta_{\text{ani}})$. We make some simplifying assumptions for this work. For the lens model parameters, $\xi_{\text{lens}}$, we assume perfect knowledge of the lens light parameters, since the uncertainty on the mass parameters and source position are dominant. We assume perfect knowledge of the image positions in the lens plane, $\theta_{\text{im}}$ assuming the contribution to the error budget from astrometry is sub-dominant. We also assume each lens has a spectroscopic measurement of the lens and source redshifts, $z_{\text{lens}}$,  $z_{\text{src}}$. Exploration of photometric redshifts is left for future work. These approximations reduce the integral in Equation~\ref{lklhd_with_redshift} to: 
\begin{multline}
    p(\mathcal{D}_k|\Omega,\nu) \propto \int p(d_{\text{td}} | \Omega, \lambda_{\text{int}}, \kappa_{\text{ext}}, \Delta\phi(\xi_{\text{lens}})) \times p(d_{\text{img}} | \xi_{\text{lens}}) \times
    p(d_{\text{kin}} | \Omega, \lambda_{\text{int}},\kappa_{\text{ext}},\mathcal{J}(\xi_{\text{lens}},\beta_{\text{ani}})) \\ p(d_{\text{los}} |\kappa_{\text{ext}}) \times p(\lambda_{\text{int}}, \beta_{\text{ani}} | \nu) \times p(\kappa_{\text{ext}}, \xi_{\text{lens}}) \;\; \mathrm{d}\lambda_{\text{int}}\mathrm{d}\kappa_{\text{ext}}\mathrm{d}\beta_{\text{ani}}\mathrm{d}\xi_{\text{lens}}.
\end{multline}
Next, we modify the image likelihood term, $p(d_{\text{img}}|\xi_{\text{lens}})$, using the exchange to a posterior-prior ratio derived in Section \ref{subsection:supporting_deriv} (Equation \ref{eqn:lklhd_to_posterior_prop}) : 
\begin{equation}
    p(d_{\text{img}}|\xi_{\text{lens}}) \propto \frac{p(\xi_{\text{lens}}|d_{\text{img}},\nu_{\text{int}})}{p(\xi_{\text{lens}}|\nu_{\text{int}})}
\end{equation}

 We use the same exchange on the line-of-sight constraint: 
\begin{equation}
    p(d_{\text{los}} | \kappa_{\text{ext}}) \propto \frac{p(\kappa_{\text{ext}} | d_{\text{los}}, \nu_{\text{int}})}{p(\kappa_{\text{ext}} | \nu_{\text{int}})}.
\end{equation}

Note this formulation is advantageous because it allows us to explicitly account for any ``interim'' modeling priors, $p(\xi_{\text{lens}}|\nu_{\text{int}})$ and $p(\kappa_{\text{ext}} | \nu_{\text{int}})$. Now, our likelihood is: 
\begin{multline}
    p(\mathcal{D}_k|\Omega,\nu) \propto \int p(d_{\text{td}} | \Omega, \lambda_{\text{int}}, \kappa_{\text{ext}}, \Delta\phi(\xi_{\text{lens}})) \times 
    p(d_{\text{kin}} | \Omega, \lambda_{\text{int}}, \kappa_{\text{ext}}, \mathcal{J}(\xi_{\text{lens}},\beta_{\text{ani}}))) \times p(\xi_{\text{lens}}|d_{\text{img}},\nu_{\text{int}}) / p(\xi_{\text{lens}}|\nu_{\text{int}})
    \\ (p(\kappa_{\text{ext}} | d_{\text{los}}, \nu_{\text{int}}) / p(\kappa_{\text{ext}} | \nu_{\text{int}}) ) \times p(\lambda_{\text{int}}, \beta_{\text{ani}} | \nu) \times p(\kappa_{\text{ext}}, \xi_{\text{lens}}) \;\; \mathrm{d}\lambda_{\text{int}}\mathrm{d}\kappa_{\text{ext}}\mathrm{d}\beta_{\text{ani}}\mathrm{d}\xi_{\text{lens}}.
\end{multline}

In this work, we make a simplifying assumption that our interim modeling priors match the true distribution over individual lens parameters, i.e.: $p(\kappa_{\text{ext}}, \xi_{\text{lens}})$ = $p(\kappa_{\text{ext}} | \nu_{\text{int}}) p(\xi_{\text{lens}}|\nu_{\text{int}})$. This assumption is valid because we are working with de-biased, emulated posteriors (see Section \ref{subsection:image_models}). In future works, the inclusion of interim prior terms may be crucial when incorporating true posteriors from modeling routines. Canceling these terms leads to a further simplified expression:
\begin{multline}
    p(\mathcal{D}_k|\Omega,\nu) \propto \int p(d_{\text{td}} | \Omega, \lambda_{\text{int}}, \kappa_{\text{ext}}, \Delta\phi(\xi_{\text{lens}})) \times 
    p(d_{\text{kin}} | \Omega, \lambda_{\text{int}}, \kappa_{\text{ext}}, \mathcal{J}(\xi_{\text{lens}},\beta_{\text{ani}}))) \times p(\xi_{\text{lens}}|d_{\text{img}},\nu_{\text{int}})
    \\ p(\kappa_{\text{ext}} | d_{\text{los}}, \nu_{\text{int}}) \times p(\lambda_{\text{int}}, \beta_{\text{ani}} | \nu) \;\; \mathrm{d}\lambda_{\text{int}}\mathrm{d}\kappa_{\text{ext}}\mathrm{d}\beta_{\text{ani}}\mathrm{d}\xi_{\text{lens}}.
\end{multline}

Finally, we evaluate the integral by importance sampling over ($\lambda_{\text{int}}, \kappa_{\text{ext}}, \beta_{\text{ani}}, \xi_{\text{lens}}$). Evaluating an integral over the kinematic likelihood is computationally expensive, as each $\mathcal{J}(\xi_{\text{lens}},\beta_{\text{ani}})$ computation takes $\sim$2 seconds. Note that the individual mass models, $p(\xi_{\text{lens}}|d_{\text{img}},\nu_{\text{int}})$ are static. Thus, a static set of importance samples over $(\xi_{\text{lens}},\beta_{\text{ani}})$ can be pre-defined, and $\mathcal{J}(\xi_{\text{lens}},\beta_{\text{ani}})$ can be pre-computed for each sample, to keep the cosmological likelihood evaluation efficient. Samples of $\xi_{\text{lens}}$ come from the image model posteriors. We introduce an interim, static sampling distribution for $\beta_{\text{ani}}$, $p(\beta_{\text{ani}} | \nu_{\text{int}})$, whose influence is then divided out during the final likelihood evaluation. This results in a final likelihood evaluation:

\begin{align}
    \nonumber p(\mathcal{D}_k \mid \Omega, \nu) \propto 
    \frac{1}{N} \sum_{\substack{\rm
        \xi_{\text{lens}},\, \beta_{\text{ani}}, \lambda_{\text{int}}, \kappa_{\text{ext}}  \sim \\
        p(\xi_{\text{lens}} \mid d_{\text{img}}, \nu_{\text{int}}) p(\beta_{\text{ani}} | \nu_{\text{int}})
        p(\kappa_{\text{ext}} \mid d_{\text{los}}) p(\lambda_{\text{int}} \mid \nu)
    }} 
    \Bigl[ 
     & p(d_{\mathrm{td}} \mid \Omega, \lambda_{\mathrm{\text{int}}}, \kappa_{\mathrm{\text{ext}}}, \Delta\phi(\xi_{\text{lens}})) \\
     \nonumber & \times\, p(d_{\mathrm{kin}} \mid \Omega, \lambda_{\mathrm{\text{int}}}, \kappa_{\mathrm{\text{ext}}}, \mathcal{J}(\xi_{\text{lens}}, \beta_{\text{ani}})) \\
     & \times\, p(\beta_{\text{ani}} \mid \nu) / p(\beta_{\text{ani}} \mid \nu_{\text{int}}) 
    \Bigr].
\label{eqn:appendix_imp_sampling_lklhd}
\end{align}

We further expand on the time-delay and kinematic likelihood terms in Section \ref{subsection:gaussian_lklhds}.

\subsection{Gaussian Measurement Likelihoods}
\label{subsection:gaussian_lklhds}

We further examine the two pieces of the likelihood containing cosmological information: $p(d_{\text{td}} | \Omega, \lambda_{\text{int}}, \kappa_{\text{ext}}, \Delta\phi(\xi_{\text{lens}}))$ and $p(d_{\text{kin}} | \Omega, \lambda_{\text{int}}, \kappa_{\text{ext}}, \mathcal{J}(\xi_{\text{lens}}, \beta_{\text{ani}}))$. In this work, we assume the time-delay measurement can be summarized by a Gaussian $\mu_{\text{obs}}(\mathbf{\Delta t})$ and $\Sigma_{\text{obs}}(\mathbf{\Delta t})$, resulting in the likelihood evaluation:
\begin{equation}
    \mathcal{N}(\mathbf{\Delta t}(\Omega, \lambda_{\text{int}}, \kappa_{\text{ext}}, \Delta\phi(\xi_{\text{lens}})) \ | \ \mu_{\text{obs}}(\mathbf{\Delta t}),\Sigma_{\text{obs}}(\mathbf{\Delta t})),
\end{equation}
where $\mathbf{\Delta t}(\Omega, \lambda_{\text{int}}, \kappa_{\text{ext}}, \Delta\phi(\xi_{\text{lens}})))$ is given by Equation \ref{eqn:time_delay_degen}.

We also assume the kinematic measurement can be summarized by a Gaussian $\mu_{\text{obs}}(\bm{\sigma_v})$ and $\Sigma_{\text{obs}}(\bm{\sigma_v})$, resulting in the likelihood evaluation:
\begin{equation}
    \mathcal{N}(\bm{\sigma_v}(\Omega, \lambda_{\text{int}}, \kappa_{\text{ext}}, \mathcal{J}(\xi_{\text{lens}}, \beta_{\text{ani}})) \ | \ \mu_{\text{obs}}(\bm{\sigma_v}),\Sigma_{\text{obs}}(\bm{\sigma_v})),
\end{equation}
where $\bm{\sigma_v}(\Omega, \lambda_{\text{int}}, \kappa_{\text{ext}}, \mathcal{J}(\xi_{\text{lens}}, \beta_{\text{ani}})))$ is given by Equation \ref{eqn:kinematics_degen}.

\section{Redshift Configuration}
\label{appendix:redshift}

\begin{figure*}[hbt!]
\begin{center}
\subfloat[$\mu(z_{\text{src}})$=2.0]{\includegraphics[scale=0.35]{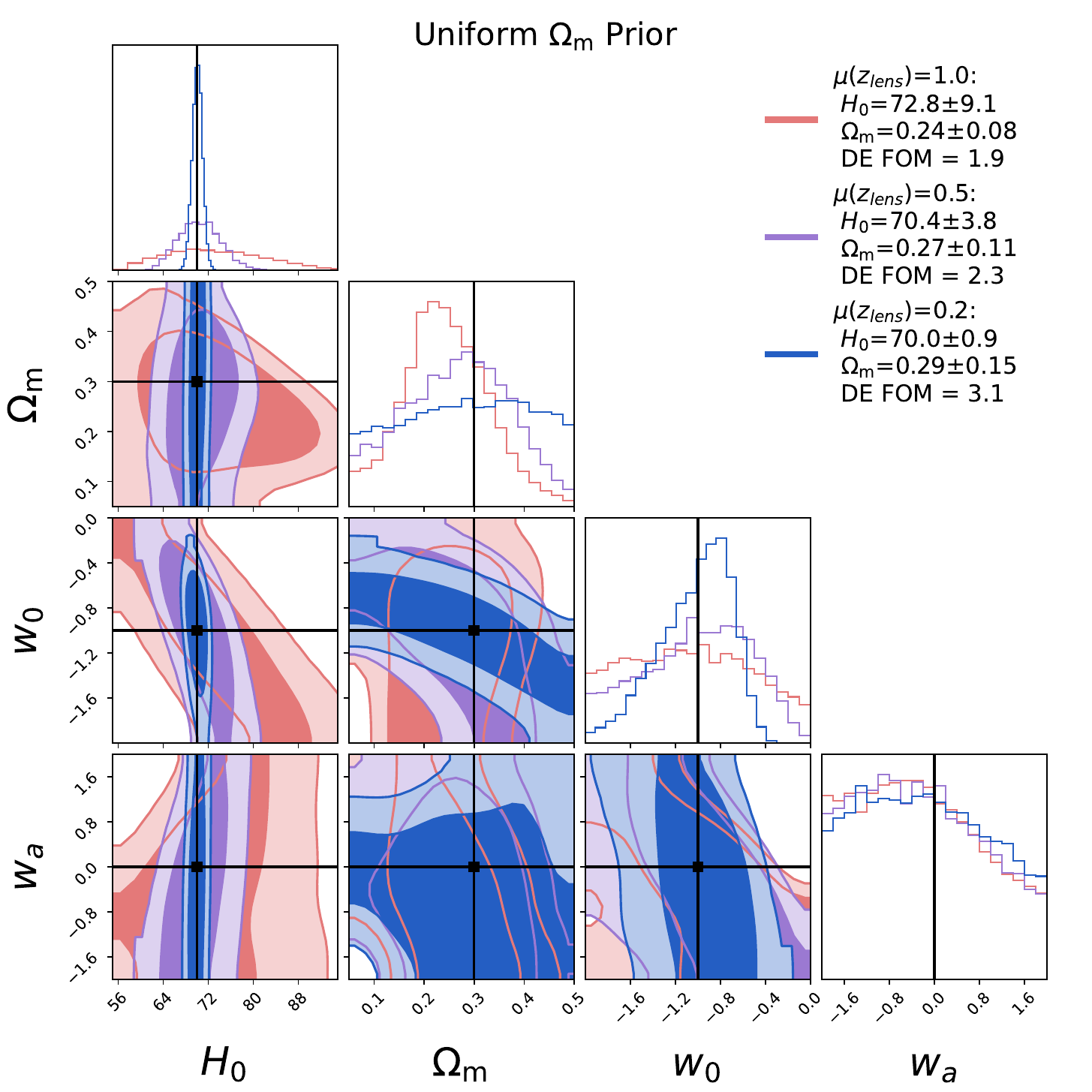}\label{fig:uni_redshift}}
\quad
\subfloat[$\mu(z_{\text{src}})$=2.0]{\includegraphics[scale=0.35]{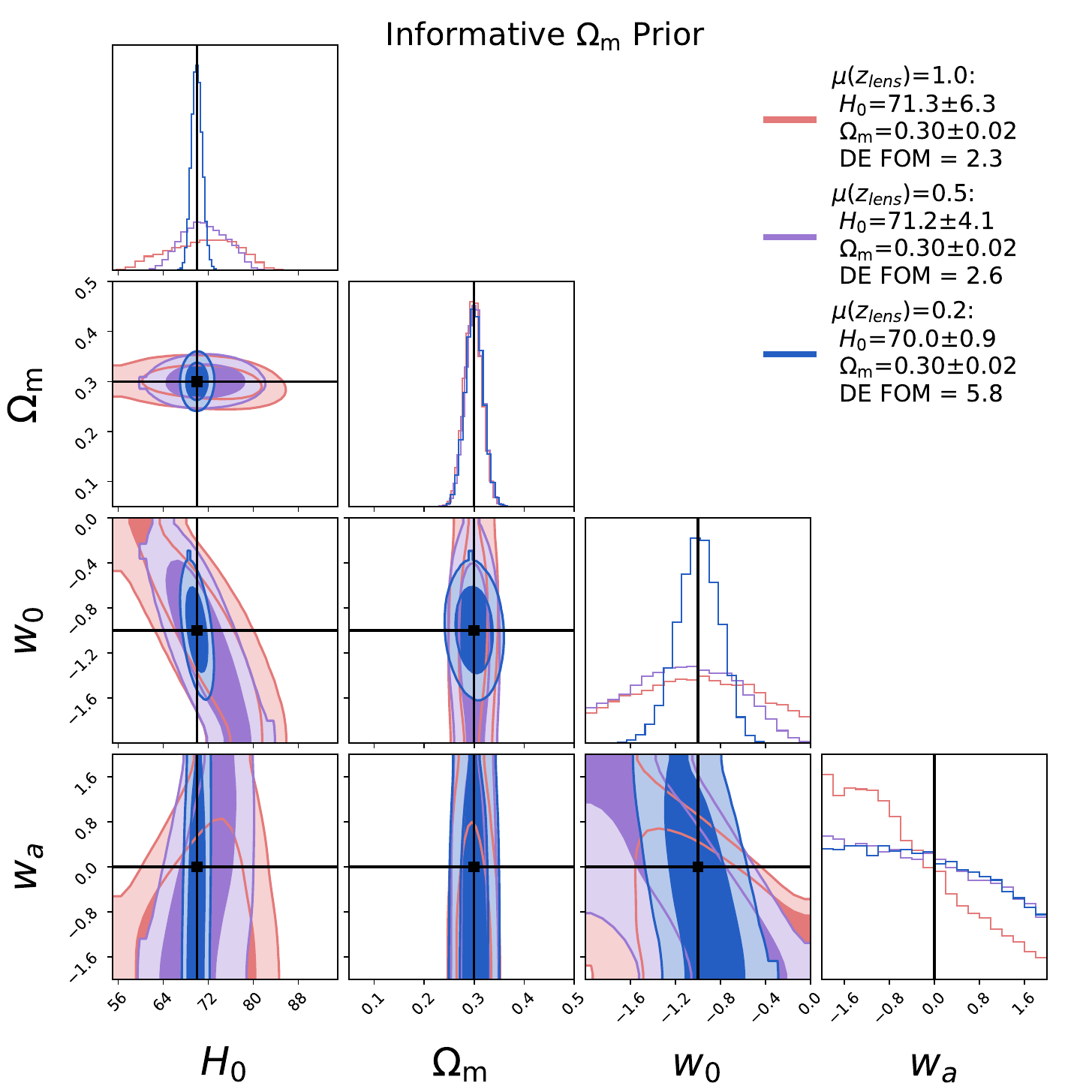}\label{fig:inf_redshift}}
\caption{Cosmological constraint from a sample of 10 lenses, changing the redshift populations. For each plot, the source redshift population is constant, with $\mu(z_{\text{src}})$=2.0, and the lens redshift changes from $\mu(z_{\text{lens}})$=1.0 (red) to $\mu(z_{\text{lens}})$=0.5 (purple) to $\mu(z_{\text{lens}})$=0.2 (blue). The left plot has a uniform $\Omega_{\text{m}}$ prior, the right plot has an informative $\Omega_{\text{m}}$ prior.}
\label{fig:redshift_contours}
\end{center}
\end{figure*}

In Section \ref{subsection:redshift_config}, we introduce a controlled test to assess the importance of the lens and source redshifts. We show select contours from the redshift configuration experiment in Figure \ref{fig:redshift_contours}. We show how changing the redshift of the lens, with a fixed source redshift at $\mu(z_{\text{src}})$ = 2, impacts the cosmological constraining power. We fix the source redshift, because as demonstrated in Table \ref{tab:redshift_exp}, the lens redshift is the dominant factor impacting cosmological constraining power, while changes to the source redshift have negligible effect. When considering the angular diameter distances in this probe, the $D_{\text{ds}}$ distance is of particular importance. When computing $D_{\text{ds}}$, changing the redshift of the deflector has a much more significant effect than changing the redshift of the source.

As discussed in Section \ref{subsection:redshift_config}, without an informative prior on $\Omega_{\text{m}}$, we see that lower lens redshift results in better constraints on $H_0$ and DE parameters, while higher lens redshift results in a better constraint on $\Omega_{\text{m}}$. When we apply an informative $\Omega_{\text{m}}$ prior, we still see a similar trend in constraining power on $H_0$ and $w_0$. Looking at the 2-Dimensional contour in ($w_0$,$w_a$), we do see at the highest lens redshift, $\mu(z_{\text{lens}})$ = 1, there appears to be an emerging ability to rule out high values of $w_a$ that is not seen at lower lens redshift. Using the DE FOM as our metric does not highlight this effect. This result may hint at the value of studying higher redshift lenses, but further investigations are required. 

We also note how changing the redshift of the lens population influences the tilt of the $H_0$-$w_0$ and $w_0$-$w_a$ contours. Strong degeneracy in the $H_0$-$w_0$ contour is also seen in \citet{Hogg_2023}. We hypothesize that the reduction in the tilt of this contour at lower lens redshift is due to a higher precision on $H_0$, which breaks this degeneracy. When designing future experiments, and considering combination with other probes, forecasting the expected tilt of the contour from time-delay cosmography will be crucial.

\section{Comparison to Previous Work}
\label{appendix:comp_to_prev}

We expand on the discussion in Section \ref{subsection:comp_to_prev_work}. We compare our analysis to the work by \citet{shajib_SLSC}, with the caveat that the two analyses make significantly different assumptions. When we compare our baseline experiment to the forecast from 236 lensed AGN in \cite{shajib_SLSC}, we see a lower DE FOM (6.7 compared to 9.6), despite our sample being almost 4x larger. We highlight two main factors.

First, we make a more conservative assumption for lenses with JWST NIRSpec kinematics. We assume ten lenses with 5\% JWST kinematics, and 40 with 5\% VLT-MUSE kinematics. In \cite{shajib_SLSC}, the sample has 40 lenses with 3\% JWST kinematics. Note from Experiment 1.1 that the spatially-resolved kinematic measurements are a significant performance driver. 

Second, we also make a more conservative assumption for the $D_{\Delta t}$ precision. We compare the 200 lenses with kinematics and space-based imaging in our sample to the 236 in \cite{shajib_SLSC}. In \cite{shajib_SLSC}, a constant precision of 5.5\% is used for every $D_{\Delta t}$. For the 50 IFU lenses in our sample, JWST-FM / HST-FM image models combined with a 3\% time-delay roughly translate to a 4-5\% constraint on $D_{\Delta t}$. However, for the 150 lenses with 4MOST kinematics, we use HST-NPE and a fixed 5-day time-delay precision. This gives a median $D_{\Delta t}$ precision around 13\%, which is significantly more conservative. 

\begin{figure*}[hbt!]
    \centering
    \includegraphics[scale=0.38]{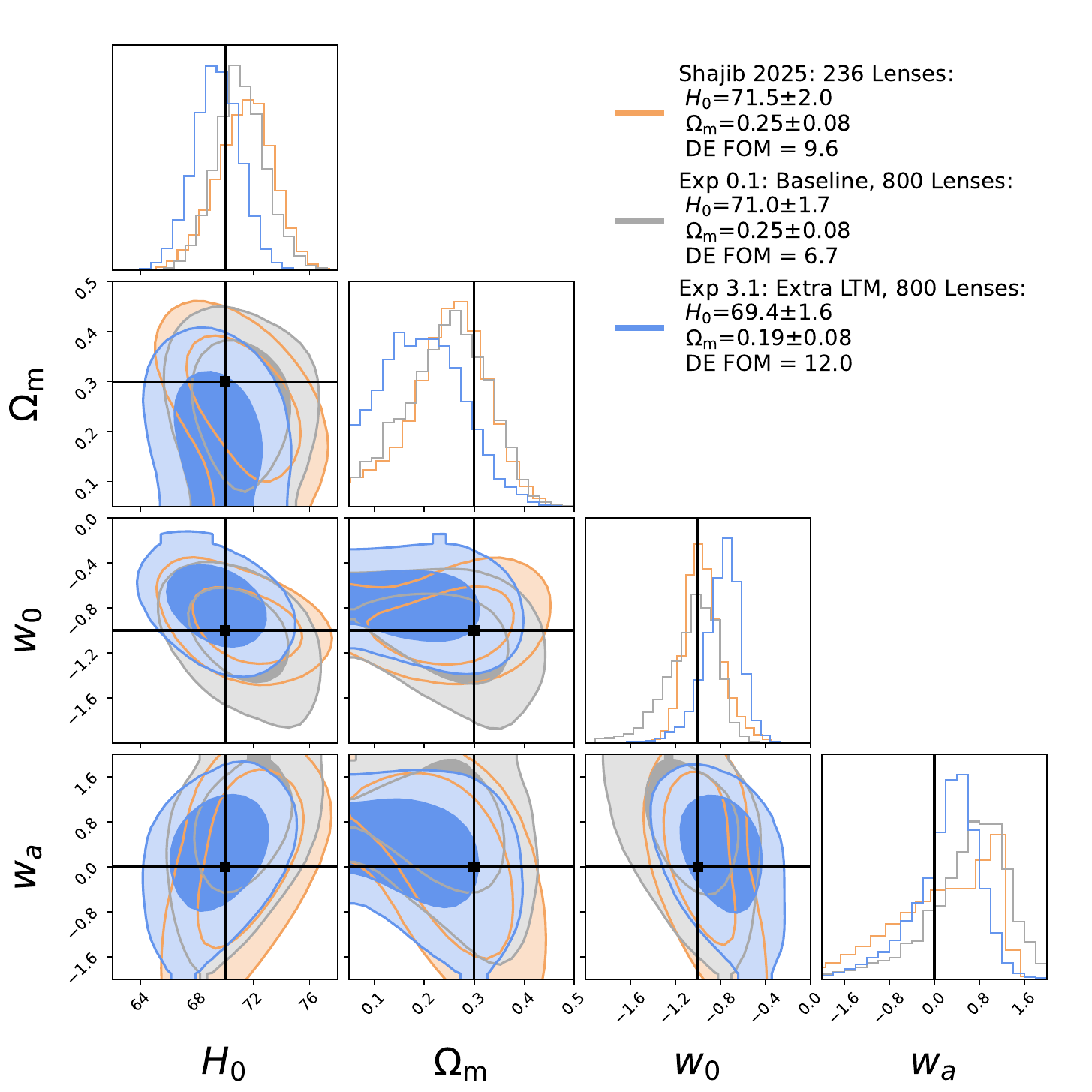}
    \caption{We compare cosmological contours from the forecast in \citet{shajib_SLSC} (orange), to our baseline experiment (grey), and our experiment 3.1 with extra long-term time-delay monitoring (blue). }
    \label{fig:comp_to_shajib}
\end{figure*}

To better match the assumptions in \cite{shajib_SLSC}, we recommend comparing to our Experiment 3.1, which upgrades the precision on the time-delays, and thus $D_{\Delta t}$, for a large portion of the single-aperture lenses. In this experiment, we do see increased constraining power compared to the \cite{shajib_SLSC} result when upgrading to the full LSST sample size (DE FOM = 12.0). We show the contours for the comparison in Figure \ref{fig:comp_to_shajib}.

Other differences include our treatment of the $\lambda_{\text{int}}$ and $\beta_{\text{ani}}$ populations, where we take a more conservative choice by assigning a non-zero scatter in both parameters. Our treatment of the kinematic measurement is also slightly different. In \cite{shajib_SLSC}, they use fully off-diagonal terms for full covariant uncertainty in the kinematic measurement, while we simplify to a diagonal measurement covariance. 

Despite using different assumptions, our work reinforces the general finding in \cite{shajib_SLSC} that there is much more constraining power available in time-delay lenses than what is shown in the DESC Science Requirements Document \citep{desc_srd}.

\section{Fluctuations with Random Seed}
\label{appendix:random_seeds}

\begin{figure*}[hbt!]
\begin{center}
\subfloat[Parameter Recovery]{\includegraphics[scale=0.5]{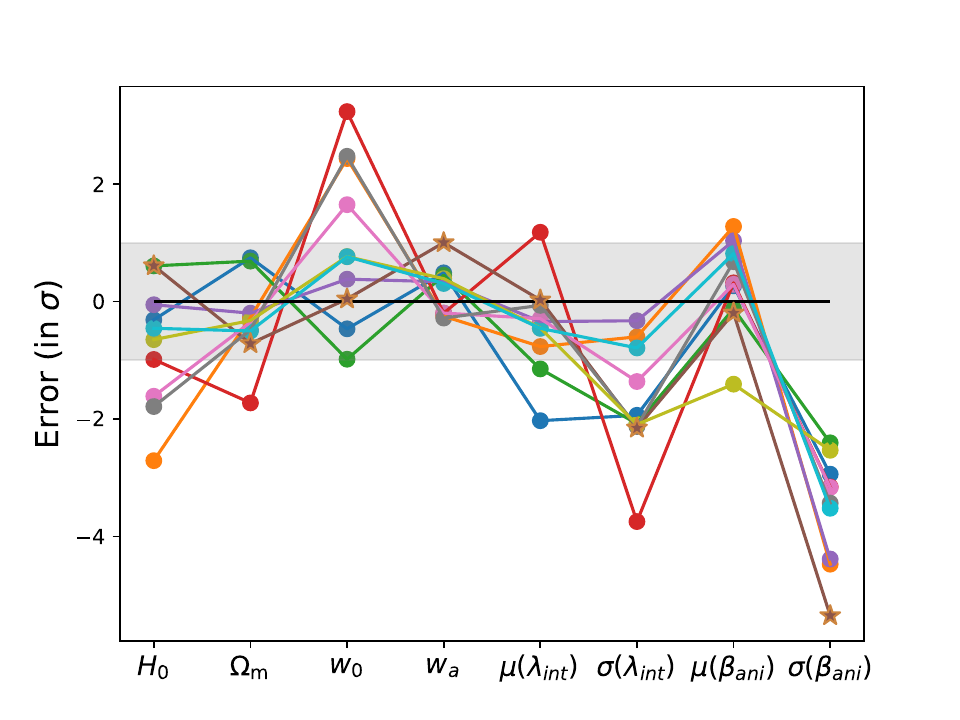}
\label{fig:scatter_all}}
\quad
\subfloat[DE FOM vs. median $w_0$]{\includegraphics[scale=0.5]{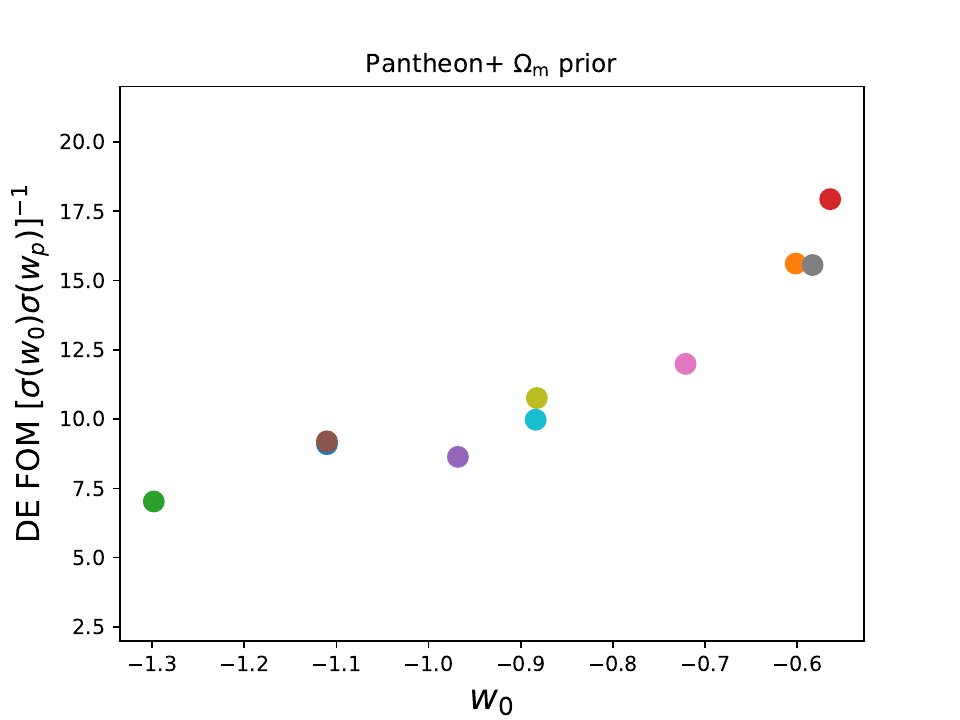}
\label{fig:scatter_fom}}
\caption{We assess the how stochasticity from measurement errors and lens selection impacts the inference by running the baseline experiment with ten random seeds. Each color corresponds to a run of the experiment with a different random seed. The baseline seed is shown in brown. \uline{On the left}, we plot the difference between the inferred value and the ground truth, divided by the 1$\sigma$ width of the posterior, for each parameter. The grey bar highlights the 1$\sigma$ region. \uline{On the right}, we show DE FOM versus the median $w_0$ value of the posterior. We see that shifts in the central value of $w_0$ correlate with the value of the DE FOM.}
\label{fig:seeds_exp}
\end{center}
\end{figure*}

We assess how the baseline results are affected by the stochasticity in lens selection and measurement error. In all previous experiments, we use a fixed random seed to control this stochasticity. Using the baseline configuration from Table \ref{table:baseline}, we run the experiment ten times, varying the random seed. We summarize the results in Figure \ref{fig:scatter_all}. As expected, results fluctuate about the ground truth values. When assessing the DE FOM across these 10 seeds, we noticed a large range of values. The DE FOM is correlated with the median values of $w_0$ and $w_a$ in each posterior, as demonstrated in Figure \ref{fig:scatter_fom}. Note that a linear change in ($w_0$,$w_a$) does \textit{not} result in a linear change in the distance-redshift relation. Assuming our constraint on the distance-redshift relation from time-delay cosmography fluctuates linearly about the ground truth due to inherent stochasticity, this results in non-linear fluctuations about the central values of $w_0$ and $w_a$. This prior volume effect on the DE FOM makes the effective precision appear higher or lower depending on the central value of the ($w_0$,$w_a$) posterior. To account for this effect, we choose a baseline seed for our experiments (brown) whose central values align with the ground truth $\Lambda$CDM values usually assumed when using the DE FOM metric. Fluctuations in preferred cosmology can impact the reported DE FOM value; this is particularly notable when comparing Experiments 3.1 and 3.4. We suggest the exploration of alternative metrics in future work.

\end{document}